\documentclass[12pt,dvips]{article}

\usepackage{rotating}
\usepackage{axodraw}
\usepackage{epsfig}
\usepackage{color}

\newcommand{\imag}{\Im {\rm m}}
\newcommand{\real}{\Re {\rm e}}

\begin{document}

\mbox{ } \\[-1cm]
\mbox{ }\hfill DESY 01--088\\
\mbox{ }\hfill IFT--01/19\\
\mbox{ }\hfill hep--ph/0108117\\
\mbox{ }\hfill \today\\

\begin{center}
{\Large{\bf ANALYSIS OF  THE NEUTRALINO SYSTEM \\[2mm]
IN SUPERSYMMETRIC THEORIES}} \\[1cm]
            S.Y. Choi$^{1,2}$,\, J. Kalinowski$^{1,3}$,\, 
            G. Moortgat--Pick$^1$ and P.M. Zerwas$^1$ 
\end{center}

\bigskip

{\small
\begin{enumerate}
\item[{}] $^1$ Deutsches Elektronen-Synchrotron DESY, D-22603 Hamburg, Germany
\item[{}] $^2$ Department of Physics, Chonbuk National University, Chonju
               561--756, Korea
\item[{}] $^3$ Instytut Fizyki Teoretycznej, Uniwersytet Warszawski, 
           PL--00681 Warsaw,     Poland
\end{enumerate}
}
\bigskip
\bigskip
\bigskip
\vskip 2cm

\begin{abstract}
  Charginos $\tilde{\chi}^\pm$ and neutralinos $\tilde{\chi}^0$ in
  supersymmetric theories can be produced copiously at 
  $e^+e^-$ colliders and their properties can be measured with
  high accuracy. Consecutively to the chargino system, in which the SU(2)
  gaugino parameter $M_2$, the higgsino mass parameter $\mu$ and
  $\tan\beta$ can be determined, 
  the remaining fundamental
  supersymmetry parameter in the gaugino/higgsino sector of the
  minimal supersymmetric extension of the Standard Model, the U(1)
  gaugino mass  $M_1$, can be analyzed
  in the neutralino system, including its modulus and its phase in
  CP--noninvariant theories. The CP properties of the neutralino
  system are characterized by unitarity quadrangles.
  Analytical solutions for the neutralino
  mass eigenvalues and for the  mixing matrix are presented for
  CP--noninvariant theories in general. They can be written in
  compact  form for large supersymmetric mass parameters.
  The closure of the neutralino and
  chargino systems can be studied by exploiting sum rules for the
  pair-production  processes in $e^+e^-$ collisions. Thus the 
  picture of the non--colored gaugino and higgsino complex 
  in supersymmetric theories can 
  comprehensively be
  reconstructed in these experiments.

\end{abstract}
%


\newpage

\section{Introduction}

In the minimal  supersymmetric extension of the Standard Model (MSSM),
the  spin-1/2 partners of the neutral gauge 
bosons, $\tilde{B}$ and $\tilde{W}^3$, 
and of the neutral Higgs bosons, $\tilde H_1^0$ and 
$\tilde H_2^0$, mix to form the neutralino mass eigenstates $\chi_i^0$
($i$=1,2,3,4). The neutralino mass matrix \cite{R1} in the 
$(\tilde{B},\tilde{W}^3,\tilde{H}^0_1,\tilde{H}^0_2)$ basis, 
\begin{eqnarray}
{\cal M}=\left(\begin{array}{cccc}
  M_1       &      0          &  -m_Z c_\beta s_W  & m_Z s_\beta s_W \\[2mm]
   0        &     M_2         &   m_Z c_\beta c_W  & -m_Z s_\beta c_W\\[2mm]
-m_Z c_\beta s_W & m_Z c_\beta c_W &       0       &     -\mu        \\[2mm]
 m_Z s_\beta s_W &-m_Z s_\beta c_W &     -\mu      &       0
                  \end{array}\right)\
\label{eq:massmatrix}
\end{eqnarray}
is built up by the fundamental supersymmetry parameters: the U(1) and
SU(2) gaugino masses $M_1$ and $M_2$, the higgsino mass parameter
$\mu$, and the ratio $\tan\beta=v_2/v_1$ of the vacuum expectation
values of the two neutral Higgs fields which break the electroweak
symmetry. Here, $s_\beta =\sin\beta$, $c_\beta=\cos\beta$ and
$s_W,c_W$ are the sine and cosine of the electroweak mixing angle
$\theta_W$.  In CP--noninvariant theories, the mass parameters are
complex.  The existence of CP--violating phases in supersymmetric
theories  in
general induces electric dipole moments (EDM).
The current experimental bounds on the EDM's can be exploited
to derive indirect limits on the parameter space \cite{CPpapers,CPmad},
which however depend on many parameters of the theory
outside the neutralino/chargino sector.

By reparametrization of the fields, $M_2$
can be taken real and positive without loss of generality so that the
two remaining non--trivial phases, which are
reparametrization--invariant, may be attributed to $M_1$
and $\mu$:
\begin{eqnarray}
M_1=|M_1|\,\,{\rm e}^{i\Phi_1}\ \qquad {\rm and} \qquad   
\mu=|\mu|\,\,{\rm e}^{i\Phi_\mu} \qquad (0\leq \Phi_1,\Phi_\mu< 2\pi)
\end{eqnarray}
The experimental analysis of neutralino properties in production and
decay mechanisms will unravel the basic structure
of the underlying supersymmetric theory.\\[2mm]
Neutralinos are produced in $e^+e^-$ collisions, either in diagonal or in
mixed pairs \cite{R2}-\cite{11a}
\begin{eqnarray*}
e^+ e^- \ \rightarrow \ \tilde{\chi}^0_i \ \tilde{\chi}^0_j 
        \qquad \ \ (i,j=1,2,3,4)
\,
\end{eqnarray*}

\noindent
If the collider energy is sufficient to produce the four neutralino
states in pairs, the underlying fundamental SUSY parameters
$\{|M_1|, \Phi_1, M_2,\,|\mu|, \Phi_{\mu}; \tan\beta\}$ can be
extracted from the masses $m_{\tilde{\chi}^0_i}$ ($i$=1,2,3,4) and
the couplings. Partial information from the lowest $m_{\tilde{\chi}^0_i}$
($i$=1,2) neutralino states \cite{kneur,CPmad,Moortgat-Pick:1999ny} is 
sufficient to
extract $\{|M_1|, \Phi_1\}$  in large parts of the parameter space 
if the other parameters have been pre--determined in the
chargino sector \cite{CDDKZ,CDGKSZ}.

The analysis will be based strictly on low--energy supersymmetry
(SUSY).  To clarify the basic structure of the neutralino sector
analytically, the reconstruction of the fundamental SUSY parameters is
carried out at the tree level; the loop corrections
\cite{loop} include parameters from other sectors of the MSSM,
demanding iterative higher--order expansions in global analyses at the
very end. When the basic SUSY parameters will have been extracted
experimentally, they may be confronted, for instance, with the
ensemble of relations predicted in Grand Unified Theories
\cite{porod}.\\[-2mm]

In this report we present a coherent and comprehensive description of
the neutralino system, discuss its properties and describe strategies
which exploit the neutralino pair production processes at $e^+e^-$ linear
colliders to reconstruct the underlying fundamental theory.  The
report is divided into six parts. In Section~2 we extend the mixing
formalism for the neutral gauginos and higgsinos to CP--noninvariant
theories with nonvanishing phases. The CP properties of the neutralino
mixing matrix are analysed in detail; 
the structure of the neutralino mixing matrix is characteristically
different from the well-known CKM and
MNS mixing matrices due to the Majorana nature of the fields involved.  
Analytic solutions for neutralino masses and
mixing matrix elements are provided for the general case, and in
compact form for the limit of large supersymmetry mass parameters
$M_{1,2}$ and $\mu$. The special toy case $M_1=M_2$ and
$\tan\beta=1$ can be solved exactly, and it illustrates 
the complex structure of CP violation in the neutralino system.  In
Section~3  the cross
sections for neutralino production with polarized beams, and the
polarization vectors of the neutralinos are given \cite{Gudi,CSS}. 
The rise of excitation curves near threshold for non--diagonal pair
production is altered qualitatively in CP--noninvariant theories.
Thus, precise measurements of the threshold behavior of the non--diagonal
neutralino pair production processes may give first indications of
non--zero CP violating phases. In Section~4 we describe the
phenomenological analysis of the complete set of the chargino and
neutralino states which  allows to extract the fundamental
SUSY parameters in a model--independent way, leading to an unambiguous
determination of the U(1) and SU(2) gaugino and higgsino parameters. 
The case in which  the analysis is 
restricted to the light neutralino states $\tilde{\chi}^0_{1,2}$
will also be discussed.  In Section~5 sum rules for the neutralino
cross sections are formulated as an experimental check of the closure
of the four-state neutralino system.  Conclusions are finally given in
Section~6.

\section{Mixing formalism}
\label{sec:mixing}

\subsection{General analysis}

In the MSSM, the four neutralinos $\tilde{\chi}^0_i$ $(i={1,2,3,4})$ are
mixtures of the neutral U(1) and SU(2) gauginos and the SU(2) higgsinos.
In the general case of CP--noninvariant theories the neutralino mass
matrix  ${\cal M}$ in eq.~(\ref{eq:massmatrix}) is complex.
Making use of possible field redefinitions, the parameters
$\tan\beta$ and $M_2$ can be chosen real and positive.  Since the
matrix ${\cal M}$ is symmetric, one 
unitary matrix $N$ is sufficient to rotate the gauge eigenstate
basis $(\tilde{B}^0,\tilde{W}^3,\tilde{H}^0_1,\tilde{H}^0_2)$ 
to the mass eigenstate basis of the Majorana fields $\tilde{\chi}^0_i$
\begin{eqnarray}
{\cal M}_{diag}=N^* {\cal M} N^{\dagger}
\label{eq:mixing matrix}
\end{eqnarray}
with
\begin{eqnarray}
   \left(\begin{array}{c}
           \tilde{\chi}^0_1 \\
           \tilde{\chi}^0_2 \\
           \tilde{\chi}^0_3 \\
           \tilde{\chi}^0_4 
         \end{array}
    \right)
   =  N\, \left(\begin{array}{c}
           \tilde{B}     \\
           \tilde{W}^3   \\
           \tilde{H}^0_1 \\
           \tilde{H}^0_2 
          \end{array}
    \right)
\end{eqnarray}
The squared mass matrix ${\cal M}_{diag}{\cal M}_{diag}^\dagger 
= N^* {\cal M} {\cal M}^\dagger N^T$ is real and positive
definite. The mass eigenvalues $m_i$ $(i=1,2,3,4)$ in ${\cal
M}_{diag}$ can  be chosen
positive by a suitable definition of the unitary matrix $N$.

The most general $4\times4$ unitary matrix $N$ can be parameterized by
6 angles and 10 phases. It is convenient to factorize the matrix 
$N$ into
a diagonal Majorana--type  ${\sf M}$ and a Dirac--type ${\sf D}$
component in the   following way:
\begin{eqnarray}
N={\sf M}\,{\sf D}
\label{eq:N}
\end{eqnarray}
with the diagonal matrix 
\begin{equation}
{\sf M}= {\sf diag}\left\{{\rm e}^{i\alpha_1},\, 
                   {\rm e}^{i\alpha_2},\,
                   {\rm e}^{i\alpha_3},\,
                   {\rm e}^{i\alpha_4}\,\right\} \label{eq:Mdef} 
\qquad (0\leq \alpha_i < \pi {\rm  ~~mod~~} \pi)
\end{equation}
One overall Majorana phase is nonphysical  and, for example, 
$\alpha_1$  may be chosen to vanish. This leaves us with 15 degrees of
freedom.   
The matrix ${\sf D}$, which depends
on 6 angles and the remaining 6 phases in four dimensions, can be written
as a sequence of 6 two-dimensional rotations \cite{sixrot}
\begin{eqnarray}
{\sf D}={\sf R}_{34}\, {\sf R}_{24}\,{\sf R}_{14}\,{\sf R}_{23}\,{\sf
  R}_{13}\, {\sf R}_{12} 
\label{eq:sixR}
\end{eqnarray}
where, for example,  
\begin{eqnarray}
{\sf R}_{12}=\left(\begin{array}{cccc}
             c_{12}  &  s^*_{12}  &  0  &  0 \\
            -s_{12}  &  c_{12}    &  0  &  0 \\
                0    &      0     &  1  &  0  \\
                0    &      0     &  0  &  1 
                  \end{array}\right)
\end{eqnarray}
The other matrices ${\sf R}_{jk}$ are defined similarly for rotations in 
the [$jk$] plane, where  
\begin{eqnarray}
&& c_{jk}\equiv \cos\theta_{jk} \qquad
s_{jk}\equiv \sin\theta_{jk}\, {\rm e}^{i\delta_{jk}}
\label{deltaphases}\\[1mm]
&&0\leq \theta_{jk} \le \pi/2 \qquad 0\leq \delta_{jk}< 2\pi 
\nonumber
\end{eqnarray}
Due to the Majorana nature of the  neutralinos, all nine phases
of the mixing matrix $N$ 
are fixed by underlying SUSY parameters,  and they cannot be removed by
rephasing the fields. CP is conserved if $\delta_{ij}=0$ or $\pi$ and
$\alpha_i=0$ mod $\pi/2$ \footnote{Majorana phases  $\alpha_i=\pm
\pi/2$  describe 
different CP parities of the neutralino states.}, {\it i.e.} the necessary 
condition for CP--noninvariance is the
non--vanishing  of at least one of the nine physical phases.

The unitary matrix $N$ of eq.~(\ref{eq:mixing matrix}) defines the
couplings of the mass eigenstates $\tilde{\chi}^0_i$ to other particles.
For the neutralino production processes it is  sufficient to
consider the  neutralino--neutralino--$Z$ vertices,
\begin{eqnarray}
&& \langle\tilde{\chi}^0_{iL}\, |Z|\tilde{\chi}^0_{jL}\rangle 
  = -\frac{g}{2\,c_W} \left[N_{i3}N^*_{j3}-N_{i4}N^*_{j4}\right]
   \nonumber\\
&& \langle\tilde{\chi}^0_{iR}\,|Z|\tilde{\chi}^0_{jR}\rangle 
  = +\frac{g}{2\,c_W} \left[N^*_{i3}N_{j3}-N^*_{i4}N_{j4}\right]
\label{eq:zvertex}
\end{eqnarray}
and the electron--selectron--neutralino vertices,
\begin{eqnarray}
&& \langle\tilde{\chi}^0_{iR}|\tilde{e}_L|e^-_L\rangle 
  = +\frac{g_{_{\tilde{W}}}}{\sqrt{2}\,c_W}\,
     \left[ N^*_{i2}\,c_W+N^*_{i1}\,s_W\right]
   \nonumber\\
&& \langle\tilde{\chi}^0_{iL}|\tilde{e}_R|e^-_R\rangle 
  = -\sqrt{2}\, g_{_{\!\tilde{B}}}N_{i1}
\label{eq:evertex}
\end{eqnarray}
The couplings $g$,  $g_{_{\tilde{W}}}$
and  $g_{_{\!\tilde{B}}}$ 
are the $We\nu $ gauge coupling, and the $\tilde{W}e\tilde{e}_L$ and
 $\tilde{B}e\tilde{e}_R$ SUSY Yukawa couplings, respectively. 
The Yukawa couplings must be identical with the SU(2) and U(1) gauge 
couplings $g$ and $g'$ at the tree level in theories in which
supersymmetry is broken softly:
\begin{eqnarray}
g_{_{\tilde{W}}}=g={e}/{s_W}\ \ {\rm and}\ \
g_{_{\!\tilde{B}}}=g'={e}/{c_W}
\end{eqnarray}
In eq.~(\ref{eq:evertex}) the coupling to the higgsino component, which
is proportional to the electron mass, has been neglected. As a result,
in the selectron vertices the $R$-type selectron couples only to
right--handed electrons while the $L$-type selectron couples only to
left--handed electrons.

\subsection{The neutralino quadrangles}

The unitarity constraints on the elements of the mixing matrix $N$
for Majorana fermions 
will first be derived without reference to the explicit form of the
neutralino mass matrix. They can be formulated by means of
unitarity quadrangles which are built up by the links $N_{ik}N^*_{jk}$  
connecting two rows $i$ and $j$,  
\begin{equation}
M_{ij}= N_{i1}N^*_{j1}+N_{i2} N^*_{j2} + N_{i3} N^*_{j3}
               +N_{i4}N^*_{j4}=0 \qquad {\rm for} \ \ i\neq j
 \label{eq:M}
\end{equation}
and by the links $N_{ki}N^*_{kj}$ connecting two  columns $i$ and $j$  
\begin{equation}
D_{ij}= N_{1i}N^*_{1j}+N_{2i} N^*_{2j} + N_{3i} N^*_{3j}
               +N_{4i}N^*_{4j}=0 \qquad {\rm for} \ \ i\neq j
 \label{eq:D}
\end{equation}
of the mixing matrix\footnote{The quadrangles $M_{ij}$ and $D_{ij}$,
when drawn in the ordering of eqs.(\ref{eq:M},\ref{eq:D}), are assumed
to be convex. Otherwise, the quadrangles can be rendered convex by 
appropriate reordering of the sides.}. There are six quadrangles of
each type.  The $M_{ij}$ quadrangles depend on the differences
of phases $\alpha_i-\alpha_j$, while the $D$--type quadrangles are not
sensitive to $\alpha_i$ phases\footnote{Corresponding to 15 degrees of
freedom, two quadrangles plus two sides and the angle in between of a
third quadrangle are independent characteristics.}.  The areas of the
six quadrangles $M_{ij}$ and $D_{ij}$ are given by
\begin{eqnarray}
{\rm area}[M_{ij}]&=&\frac{1}{4}
         (|J_{ij}^{12}|+|J_{ij}^{23}|+|J_{ij}^{34}|+|J_{ij}^{41}|)
\label{eq:aM}\\
{\rm area}[D_{ij}]&=&\frac{1}{4}
         (|J_{12}^{ij}|+|J_{23}^{ij}|+|J_{34}^{ij}|+|J_{41}^{ij}|)
\label{eq:aD}
\end{eqnarray}
where $J_{ij}^{kl}$ are the Jarlskog--type CP--odd ``plaquettes''
\cite{CJ} 
\begin{equation}
J_{ij}^{kl}=\imag N_{ik}N_{jl}N_{jk}^*N_{il}^*
\label{eq:plaq}
\end{equation}
The plaquettes are insensitive to the $\alpha_i$ phases. There are
nine independent plaquettes \cite{nine}, for example $J_{12}^{12}$, 
$J_{12}^{23}$,  $J_{12}^{34}$, $J_{13}^{12}$, $J_{13}^{23}$, $J_{13}^{34}$,
$J_{23}^{12}$, $J_{23}^{23}$, $J_{23}^{34}$. 
If they all are zero, all other plaquettes are also zero. The matrix
$N$ is CP violating, if either any one of the plaquettes is non--zero, or, 
if the plaquettes all vanish, at least one of the links is non--parallel to
the real or to the imaginary axis.

Since the phases of the neutralino fields are fixed (modulo a common
phase), the orientation of the neutralino quadrangles  $M_{ij}$ and
$D_{ij}$  in the complex plane is physically meaningful. 
This is in contrast to the CKM unitarity triangles which
all can be rotated by rephasing the left--chiral quark fields; in the
4--family case only three $\delta$ (Dirac) phases would therefore be
physical.  It is also in contrast to the $D$--type MNS
unitarity triangles  which can be rotated by rephasing the
left--chiral charged--lepton fields while, on the other hand, the
orientation of the $M$--type triangles is fixed by the phases of the
neutrino Majorana fields; in the 4--family case, three $\alpha$
(Majorana) and three $\delta$ (Dirac) phases would be observables.

\begin{figure}[htb]
\begin{center}
  \epsfig{file=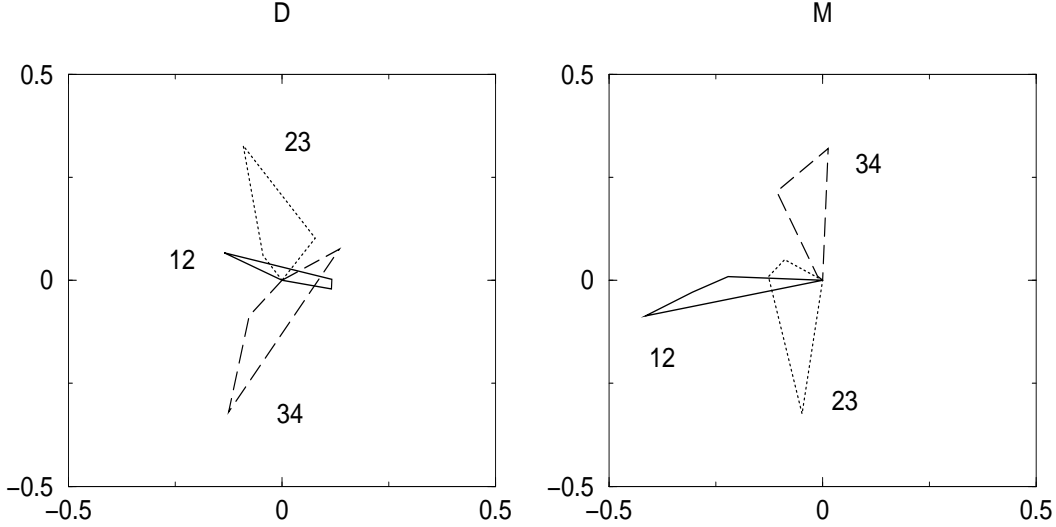,height=14cm,width=7cm,angle=-90} 
  \vspace*{1.5cm}
\caption{\it The $D$--type (left panel) and $M$--type (right panel) 
  quadrangles in the complex plane,
  illustrated for $\tan\beta=3$, $|M_1|=100$ GeV, $\Phi_1=0$,
  $M_2=150$ GeV, $|\mu|=200$ GeV and $\Phi_{\mu}=\pi/2$; 
  $ij$ as indicated in the figure.}
\label{quadreal}
\end{center}
\end{figure}

In Fig.~\ref{quadreal} two sets of three (independent) quadrangles of
each type ($M_{12}$, $M_{23}$,
$M_{34}$, and $D_{12}$, $D_{23}$, $D_{34}$) are shown for
illustration. 
The collapsing of three
quadrangles in one set (for instance $M_{12}$, $M_{23}$ and $M_{34}$)
would imply the vanishing of all plaquettes and, consequently, the areas
of all quadrangles would be zero.  However, this does {\it not} imply
the vanishing of all $\delta$-type phases (to be contrasted to the CKM and
MNS cases, where the vanishing areas of three independent quadrangles
implies the vanishing of all Dirac phases \cite{delAguila:1996pa}), as
demonstrated explicitly in Fig.\ref{quaddeg} for a special case.
Since the orientation of both $M$- and $D$-type quadrangles is
non--trivial, CP is conserved in the neutralino system only if all
quadrangles have null areas {\it and} if they all collapse to lines
oriented along the real or the imaginary axis.

By measuring only the amplitudes for neutralino pair production in
$e^+e^-$ collisions, the links of the quadrangles $M_{ij}$ and
$D_{ij}$ cannot be reconstructed completely. The relevant interactions
involving (nearly zero--mass) electron fields are invariant under the
chiral rotations,
\begin{eqnarray}
&& \tilde{H}^0_1\, \rightarrow\, {\rm e}^{i\theta_1 \gamma_5}\, \tilde{H}^0_1
\qquad \qquad
\tilde{H}^0_2\, \rightarrow\, {\rm e}^{i\theta_2\gamma_5}\, \tilde{H}^0_2
        \nonumber\\[2mm]
&&\;\, \tilde{B}  \,  \rightarrow\, {\rm e}^{i\theta_3\gamma_5}\, \tilde{B} 
\qquad \qquad\;\;\,
\tilde{W}^3\,  \rightarrow\, {\rm e}^{i\theta_3\gamma_5}\,  \tilde{W}^3
\label{eq:rephasing}
\end{eqnarray}
applied to  the weak eigenstates.  The higgsino fields can be
redefined with different phases, leaving the
$Z$--neutralino--neutralino vertices unchanged, eq.~(\ref{eq:zvertex}).
On the other hand, the electron--selectron--neutralino interaction
vertices, eq.~(\ref{eq:evertex}), are invariant under the redefinition
of the SU(2) and U(1) gaugino fields, $\tilde{W}^3$ and $\tilde{B}$,
only with an identical phase due to the non--trivial mixing of the two
gaugino states after electroweak gauge symmetry breaking.  All these
chiral phase rotations give rise to the same neutralino mass spectrum.  Under
the rephasing in eq.~(\ref{eq:rephasing}), five of the $D$-type
quadrangles rotate in the complex plane, while the orientation of
$D_{12}$ and of all $M_{ij}$ quadrangles is fixed.  As a result, out
of nine phases three of the $\delta$-type phases remain ineffective,
leaving only six phases which can be determined from $e^+e^-$
production processes: three of the $\alpha$-type and three of the
$\delta$-type.

Thus the neutralino production processes alone do not allow to 
reconstruct {\it all} the links 
of the quadrangles ${M}_{{ij}}$ and ${D}_{{ij}}$. 
However,  if interactions involving other fermion--sfermion--neutralino 
vertices of left--handed sfermions are taken into account, at least the 
$M$--type quadrangles $M_{ij}$ can be reconstructed in total, because 
the new vertices probe different combinations of the bino and wino 
components of the neutralino:
\begin{eqnarray}
&& \langle\tilde{\chi}^0_{iR}|\tilde{f}_L|f_L\rangle 
  = -\sqrt{2}\,\,\frac{g_{_{\tilde{W}}}}{c_W}\,
     \left[\,T_{3L}^f N^*_{i2}\,c_W+(Q_f-T_{3L}^f) N^*_{i1}\,s_W\,\right]
\label{eq:new vertex}
\end{eqnarray}
For example, $N_{i1}N^*_{j1}$ and $N_{i2}N^*_{j2}$ as well
as $\real(N_{i1}N^*_{j2})$ can be disentangled from two
electron--selectron--neutralino   
and one neutrino--sneutrino--neutralino interaction. Exploiting
subsequently the unitarity condition ${M}_{ij}=\delta_{ij}$,
eq.~(\ref{eq:M}),   and the 
$Z \tilde{\chi}^0_i \tilde{\chi}^0_j$ interactions, the four
sides of the quadrangle $M_{ij}$ can be determined completely.

Since the neutralino mass matrix involves only two
invariant phases $\Phi_1$ and $\Phi_\mu$, all the physical
phases of $N$ are fully determined by these two phases in the mass
matrix as well as by the gaugino/higgsino masses and the mixing parameter
$\tan\beta$. In this context, the
measurement of the $\alpha$ and the $\delta$ phases and the experimental
reconstruction of the unitarity quadrangles overconstrains the
neutralino system  and
numerous consistency relations can be exploited to scrutinize the
validity of the underlying theory. \\

\subsection{Neutralino masses and mixing matrix: analytical solutions}

Complete analytical solutions can be derived for the neutralino mass
eigenvalues $m_i\equiv m_{\tilde{\chi}^0_i}>0$ ($i=1,\ldots,4$) and
for the mixing matrix $N$ as functions of the SUSY parameters $\{|M_1|,
\Phi_{1}, M_2, |\mu|, \Phi_{\mu};\tan\beta\}$.  While earlier analyses in
Ref.\cite{BEG} were restricted to a CP--invariant neutralino sector,
we extend the analysis to the more general case of CP--violating 
theories.

For this purpose switching to  the  basis 
$(\tilde{\gamma},\tilde{Z}^0,\tilde{H}^0_a,\tilde{H}^0_b)$ by the
transformation 
\begin{eqnarray}
    \left(\begin{array}{c}
           \tilde{\gamma} \\
           \tilde{Z}^0 \\
           \tilde{H}^0_a \\
           \tilde{H}^0_b 
         \end{array}
    \right)   = 
{\cal A}\,  \left(\begin{array}{c}
           \tilde{B}     \\
           \tilde{W}   \\
           \tilde{H}^0_1 \\
           \tilde{H}^0_2 
          \end{array}
    \right) = 
\left(\begin{array}{cccc}
             c_W  &  s_W  &  0  &  0 \\
            -s_W  &  c_W    &  0  &  0 \\
                0    &      0     &  c_\beta  &  -s_\beta  \\
                0    &      0     &  s_\beta  &  c_\beta 
                  \end{array}\right)\;
 \left(\begin{array}{c}
           \tilde{B}     \\
           \tilde{W}   \\
           \tilde{H}^0_1 \\
           \tilde{H}^0_2 
          \end{array}
    \right)
\label{eq:mixC}
\end{eqnarray}
is of great advantage.
In this basis the mass matrix $\hat{\cal M}$ takes the form
\begin{eqnarray}
\hat{\cal M}= {\cal A}{\cal M} {\cal A}^T=  \left(\begin{array}{cccc}
  M_1 c^2_W+M_2 s^2_W & (M_2-M_1)\, s_W c_W & 0  & 0 \\[2mm]
  (M_2-M_1)\, s_W c_W  & M_1 s^2_W+M_2\, c^2_W &   m_Z  & 0\\[2mm]
0 & m_Z &       \mu s_{2 \beta}       &     -\mu c_{2 \beta}    \\[2mm]
0 & 0 &     -\mu c_{2 \beta}      &       -\mu s_{2 \beta}
\end{array}\right)\
\label{eq:massmatrixg}
\end{eqnarray}
where $M_1$ and $\mu$ are complex--valued; $s_{2\beta}=\sin 2\beta$ and
$c_{2\beta}=\cos 2\beta$.  The transformation ${\cal A}$ shifts zeros 
in the diagonal of ${\cal M}$ to the
non--diagonal elements of $\hat{\cal M}$ which simplifies the
solution of the eigenvalue equation (\ref{eq:eigeneq}) considerably.

The unitary matrix $\hat N$ diagonalizing the mass matrix $\hat{\cal 
M} \rightarrow {\cal M}_{diag}$ may be decomposed into the Majorana part
${\sf M}$, equivalent to eq.~(\ref{eq:N}), and the $\hat{\sf D}$ part
as follows:
\begin{eqnarray}
\hat N={{\sf M}}{\hat{\sf D}}
\label{eq:Nhat}
\end{eqnarray}
The two unitary transformations are connected by $N=\hat N {\cal A}$. The
square of the diagonal matrix ${\cal M}_{diag}$ is related to $\hat{\cal
  M}$ by the transformation
\begin{equation} 
{\cal M}_{diag}{\cal M}_{diag}^\dagger
= \hat{\sf D}^* \hat{\cal M} \hat{\cal M}^\dagger \hat{\sf D}^T 
\label{eq:masseig}
\end{equation}
The diagonal mass matrix ${\cal M}_{diag}$ itself can be defined by
the {\it positive} diagonal elements
\begin{equation}
{\cal M}_{diag}= {\sf diag}\left\{m_1, m_2, m_3, m_4\right\} >0
\end{equation} 
choosing suitable solutions for  the  phases  $\alpha_i$ in  the
matrix ${\sf M}$ derived from the  equation
\begin{eqnarray} 
{\sf M}^2 {\cal M}_{diag}&=&\hat{\sf D}^* \hat{\cal M} \hat{\sf D}^{-1}
\label{eq:phasM}
\end{eqnarray}

The mass eigenvalues $m_i^2$ ($i$=1,2,3,4), not necessarily ordered yet 
in the sequence of increasing values, are derived from
eq.~(\ref{eq:masseig}) rewritten as the eigenvalue equation
\begin{equation}
[\hat{\cal M} \hat{\cal M}^\dagger - m_i^2] \hat{\sf D}_i =0
\label{eq:eigeneq}
\end{equation}
where the eigenvectors $ \hat{\sf D}_i=(\hat{\sf D}_{i1}, \hat{\sf
D}_{i2}, \hat{\sf D}_{i3}, \hat{\sf D}_{i4})$ denote the rows of the
unitary matrix $\hat{\sf D}$.  The eigenvalues $m_i^2$ are the
solutions of the characteristic equation
\begin{equation}
m_i^8-a m_i^6+b m_i^4-c m_i^2+d=0
\label{eq:characteristic}
\end{equation}
with the invariants $a$, $b$, $c$ and $d$ given\footnote{{\it Post
  festum}  the invariants can also be
  rewritten in terms of the mass eigenstates:\\ 
  $a=m_1^2+m_2^2+m_3^2+m_4^2$\\ $b=m_1^2m_2^2+m_1^2m_3^2+m_1^2 m_4^2
  +m_2^2m_3^2 +m_2^2m_4^2 +m_3^2m_4^2$\\ $c=m_1^2m_2^2m_3^2 +
  m_1^2m_2^2m_4^2 + m_1^2m_3^2m_4^2 + m_2^2m_3^2m_4^2$\\
  $d=m_1^2m_2^2m_3^2m_4^2$. }
 by the fundamental
parameters of the neutralino system in ${\cal X}={\cal M}
{\cal M}^\dagger$:
\begin{eqnarray}
&& a = {\sf tr}{\cal X} \nonumber \\
&& \hskip 0.3cm = |M_1|^2+M^2_2+2 |\mu|^2+ 2m^2_Z
\nonumber \\[3mm]
&& b= \frac{1}{2}\left[({\sf tr}{\cal X})^2 
                           - {\sf tr}{\cal X}^2\right]\nonumber\\
&& \hskip 0.3cm = |M_1|^2 M^2_2+2|\mu|^2(|M_1|^2+ M^2_2)+(|\mu|^2+m_Z^2)^2
     \nonumber\\
&& \hskip 0.8cm + 2m^2_Z\, \{|M_1|^2 c_W^2+M_2^2s_W^2-s_{2\beta}|\mu| 
           (|M_1|s^2_W\cos(\Phi_1+\Phi_\mu)
     +M_2 c^2_W \cos\Phi_\mu)\, \}\nonumber\\[3mm]
&& c= \frac{1}{6}\left[({\sf tr}{\cal X})^3
                           -3{\sf tr}{\cal X} {\sf tr}{\cal X}^2
                           +2{\sf tr}{\cal X}^3\right]\nonumber\\
&& \hskip 0.3cm = |\mu|^2\, \{|\mu|^2 (|M_1|^2+M_2^2)+2 |M_1|^2M_2^2 +  
m_Z^4 s^2_{2\beta} +2m_Z^2(|M_1|^2c_W^2+M_2^2s_W^2)\, \} \nonumber \\
&& \hskip 0.7cm -2 m_Z^2 |\mu| s_{2\beta}\, \{|M_1|(|\mu|^2+M_2^2)s_W^2
     \cos(\Phi_1+\Phi_\mu)
+M_2(|\mu|^2+|M_1|^2)c_W^2\cos\Phi_\mu\, \}\nonumber \\
&& \hskip 0.8cm  +m_Z^4\,\{|M_1|^2c_W^4+2 |M_1|M_2s_W^2 c_W^2 \cos\Phi_1 
 +M_2^2 s_W^4\,\} \nonumber\\[3mm]
&& d = {\sf det}{\cal X}\nonumber \\
&& \hskip 0.3cm 
= |\mu|^4 M_2^2 |M_1|^2-2m_Z^2 |\mu|^3 |M_1| M_2 s_{2\beta} \,\{ 
|M_1| c_W^2 \cos\Phi_\mu + M_2 s_W^2\cos(\Phi_1+\Phi_\mu)\,\}\nonumber \\ 
&& \hskip 0.8cm  +m_Z^4 |\mu|^2 s^2_{2 \beta}
\,\{ |M_1|^2c_W^4+2 |M_1|M_2 s_W^2c_W^2
\cos\Phi_1 +M_2^2 s_W^4\,\}
\label{eq:inv}
\end{eqnarray}
Using standard methods for the solution of the quartic equation
\cite{Bron},  the eigenvalues 
\begin{eqnarray}
&& 2m_1^2 = +\sqrt{z_1}+\sqrt{z_2}-\sqrt{z_3}+ a/2
\nonumber\\
&& 2m_2^2 = +\sqrt{z_1}-\sqrt{z_2}+\sqrt{z_3}+a/2
\nonumber\\
&& 2m_3^2 = -\sqrt{z_1}+\sqrt{z_2}+\sqrt{z_3}+a/2
\nonumber\\
&& 2m_4^2 = -\sqrt{z_1}-\sqrt{z_2}-\sqrt{z_3}+a/2
\label{masses2}
\end{eqnarray}
can be expressed in terms of the roots of the triple resolvent
equation, 
\begin{eqnarray}
&& z_1 = 2\tilde{z}-{2} p/3 \nonumber\\
&& z_2 = -\tilde{z}+ \sqrt{-3 \tilde{z}^2-3 
     \tilde{p}}-{2} p/3 \nonumber\\
&& z_3 = -\tilde{z}- \sqrt{-3 \tilde{z}^2-3 
  \tilde{p}}-{2} p/3
\end{eqnarray}
with the abbreviations 
\begin{eqnarray}
&& \tilde{z} = [(-\tilde{q}+\sqrt{\tilde{q}^2+\tilde{p}^3})^{\frac{1}{3}}
  +(-\tilde{q}-\sqrt{\tilde{q}^2+\tilde{p}^3})^{\frac{1}{3}}]/2\nonumber \\
&&\tilde{p}=-{p}^2/9-{4} r/3\nonumber \\
&&\tilde{q}=-{p}^3/27+4 r {p}/{3}-{q^2}/{2}
\end{eqnarray}
which are defined by the invariants
\begin{eqnarray}
&&p= -3 {a}^2/{8}+b \nonumber \\ 
&&q= -{a}^3/8+{a}b/{2}-c \nonumber \\
&&r= -3 {a}^4/{256}+{a}^2 b/{16} -{a}c/{4} c+d
\end{eqnarray}
When taking the square roots of the $z_i$, the  signs of two roots are
arbitrary, just reordering the eigenvalues when signs are switched,
while the sign of the third root is predetermined by the Vieta 
condition  $\sqrt{z_1} \sqrt{z_2} \sqrt{z_3}=q$.

The elements of the mixing matrix $\hat{\sf D}$ follow from the
eigenvector equation (\ref{eq:eigeneq}), 
\begin{eqnarray}
 \hat{\sf D}_i = ({B_i}/{A_i}{N_i}, 
     {1}/{N_i}, {C_i}/{A_i}{N_i},  {D_i}/{N_i})
\label{eq:Dhat}
\end{eqnarray}
where
\begin{eqnarray}
&& A_i = m_Z^2 (M_2^2 s^4_W+|M_1|^2 c^4_W+2 s^2_W c^2_W M_2 |M_1|\cos\Phi_1
  -m_i^2) \nonumber\\
  & & \hskip 1.5cm +(M_2^2 s^2_W+|M_1|^2 c^2_W-m_i^2)(|\mu|^2-m_i^2)
\nonumber\\[2mm]
&& B_i =  s_W c_W [ m_Z^2(M_1 c^2_W+M_2 s^2_W)(M_1^*-M_2)
      +m_Z^2 \mu (M_2-M_1) s_{2\beta} \nonumber\\
   && \hskip 1.5cm  -(|\mu|^2-m_i^2)(M_2^2-|M_1|^2)] \nonumber\\[2mm]
&& C_i = m_Z\left[M_1^* s_W^2(m_i^2-M_2^2)+ M_2 c^2_W(m_i^2-|M_1|^2) 
      -\mu s_{2\beta} (M_2^2 s^2_W+|M_1|^2 c^2_W-m_i^2)\right]
   \nonumber\\[2mm] 
&& D_i = \frac{m_Z \mu\, c_{2\beta}}{|\mu|^2-m_i^2}
\end{eqnarray}
and the normalization condition
\begin{equation}
N_i= \left[1+(|B_i|^2+|C_i|^2)/A_i^2+ |D_i|^2\right]^{1/2}
\end{equation}
which completes the eigensystem. 

Factorizing the matrix $\hat{\sf D}$ into six 2$\times$2
rotations, as defined in  eq.~(\ref{eq:sixR}), the most compact
representation for the mixing angles $ \theta_{ij}$ 
and the phases $\delta_{ij}$ is given in terms of the sines 
$s_{ij}=\sin\theta_{ij} {\rm e}^{i\delta_{ij}}$ by 
%
\begin{eqnarray}
&& s_{12}= \frac{A_1}{[A_1^2 N^2_1 (1-|D_1|^2/N_1^2)-|C_1|^2]^{1/2}}
           \nonumber\\
&& s_{13}= \frac{C_1^*}{A_1 N_1 \sqrt{1-|D_1|^2/N_1^2}}\nonumber\\[1mm]
&& s_{14}= \frac{D_1^*}{N_1} \nonumber\\  
&& s_{23}=\frac{A_1 C_2^* N_1\, (1-|D_1|^2/N_1^2)/A_2
         +C_1^* D_1 D^*_2/N_1}{
   \left[A_1^2 N^2_1(1-|D_1|^2/N_1^2)-|C_1|^2 \right]^{1/2}
   \left[N_2^2(1-|D_1|^2/N_1^2)-|D_1|^2 \right]^{1/2}}\nonumber\\[1mm]
&&  s_{24}= \frac{D_2^*}{N_2\sqrt{1-|D_1|^2/N_1^2}}\nonumber\\
&&  s_{34}= \frac{N_2 D^*_3}{N_3[N_2^2(1-|D_1|^2/N_1^2)-|D_2|^2]^{1/2}}
\end{eqnarray}
%

%
%

The phases $\alpha_i$ in  the Majorana  matrix ${\sf M}$ are
derived from 
 \begin{eqnarray} 
 {\rm e}^{2i\alpha_i} &=& \sum_k\sum_l \hat{\sf D}^*_{ik}\, \hat{\sf D}^*_{il}
                          \, \hat{\cal M}_{kl}/m_i\nonumber\\ 
  &=& \bigg\{(B^*_i/A_i)^2 (M_1 c^2_W+M_2 s^2_W)
            + 2 (B^*_i/A_i)\, (M_2-M_1)\, s_W c_W 
	    + M_1 s^2_W + M_2 c^2_W \nonumber\\
  && \mbox{ }\hskip 0.1cm + 2 (C^*_i/A_i) m_Z 
     +\left[(C^*_i/A_i)^2 - D^{*2}_i\right] \mu s_{2\beta}
     -2(C^*_i/A_i) D^*_i \mu \, c_{2\beta} \bigg\}/m_i 
 \label{eq:phasD}
 \end{eqnarray}
with positively chosen eigenvalues $m_i>0$ in ${\cal M}_{diag}$, 
and the matrix elements given in
 eq.~(\ref{eq:Dhat}). The 
$\alpha_i$  can finally be reparametrized such that $\alpha_1=0$ and
$0\leq \alpha_{2,3,4} < \pi$ in general.

\subsection{Compact solutions in special cases}
A particularly interesting limit is approached when the supersymmetric
mass parameters (and their splittings) 
are considerably larger than the electroweak scale:  $ M_{SUSY}^2\gg
m_Z^2$. In 
this limit a compact approximate solution for the neutralino masses and mixing
angles can be derived. 
On the other hand, in the special case 
of gaugino mass degeneracy
$M_1=M_2$ in the limit $\tan\beta=1$, the exact solutions for the mass
eigenvalues and the mixing matrix can be presented in a compact closed
form. 
Though somewhat academic, this  configuration will allow us  to
illustrate some surprising consequences of CP--violation for the
structure of the neutralino
sector in a very transparent way.

\subsubsection{The mixing matrix at large SUSY scales}
If the supersymmetry mass parameters, $M_{1,2}^2$ and $|\mu|^2$, and their
splittings are much larger than $m_Z^2$, $|M_{1,2}|^2, \, |\mu|^2 \gg
m_Z^2$  and
$\left||M_{1,2}|\pm|\mu|\right|^2\gg m_Z^2$,  
the diagonalization of the 
neutralino mass matrix can be expanded 
in the two small (dimensionless) parameters
\begin{eqnarray}
 X_1 =\frac{m^2_Z\, s^2_W}{|M_1|^2-|\mu|^2} \qquad {\rm and} \qquad
 X_2 =\frac{m^2_Z\, c^2_W}{|M_2|^2-|\mu|^2}
\end{eqnarray}
The corresponding expansion in the CP--conserving case for both
charginos and neutralinos had been worked out in Ref.~\cite{oldexp}; 
we generalize this expansion by including arbitrary phases.

In the limit of large SUSY scales the mixing matrix $N$ can be cast into 
a compact form by factorizing the matrix in yet another form as follows:
\begin{eqnarray}
N= {\sf M}\,  {\sf D}'\, {\sf P} \label{eq:approxN}
\end{eqnarray}
where the unitary matrix ${\sf D}'$ is isomorphic to the form 
given in eq.~(\ref{eq:sixR}) with redefined sines and cosines due to
the presence of ${\sf P}$. This matrix is
conveniently chosen as
\begin{eqnarray}
&& {\sf P}={\sf diag}\left\{{\rm e}^{\frac{i}{2}\Phi_1},  \, 
                           1,                  \,
            {\rm e}^{\frac{i}{2}\Phi_\mu},\,
            {\rm e}^{\frac{i}{2}\Phi_\mu}\,\right\}
            \left(\begin{array}{cccc}
             1    &  0   &         0           &         0 \\
             0    &  1   &         0           &         0 \\
             0    &  0   & \frac{1}{\sqrt{2}}  & -\frac{1}{\sqrt{2}} \\
             0    &  0   & \frac{1}{\sqrt{2}}  &  \frac{1}{\sqrt{2}} 
                  \end{array}\right)
\label{eq:diagonal}
\end{eqnarray}

Retaining the leading  order  in $X_1$ and $X_2$, the neutralino mass  
eigenvalues (not ordered yet sequentially with increasing
mass) are given as 
\begin{eqnarray}
&& m_1 = |M_1| + X_1\bigg[|M_1|+|\mu|\cos 2\eta
         \cos(\Phi_1+\Phi_\mu)\bigg]\nonumber\\[1mm]
&& m_2 = |M_2| + X_2\bigg[|M_2|+|\mu|\cos 2\eta
         \cos \Phi_\mu\bigg]\nonumber\\[1mm]
&& m_3 = |\mu|- c^2_\eta\, \bigg[ (X_1+X_2)|\mu|
        + X_1|M_1|\cos(\Phi_1+\Phi_\mu)
        + X_2|M_2|\cos\Phi_\mu \bigg]\nonumber\\[1mm]
&& m_4 = |\mu|- s^2_\eta\, \bigg[ (X_1+X_2)|\mu|
        - X_1|M_1|\cos(\Phi_1+\Phi_\mu)
        - X_2|M_2|\cos\Phi_\mu \bigg]
\end{eqnarray}
where 
$ c_\eta=(c_\beta+s_\beta)/\sqrt{2}$ and 
$s_\eta=(c_\beta-s_\beta)/\sqrt{2}$. 
The unitary matrix ${\sf D}'$ is approximately represented by 
\begin{eqnarray}
{\sf D}' = \left(\begin{array}{cccc}
      c_{13}c_{14}  &   s^*_{12}    & s^*_{13}      & s^*_{14}   \\[1mm]
      -{s'}_{12} & c_{23}c_{24}  & s^*_{23}      & s^*_{24}   \\[1mm]
        -s_{13}     &  -s_{23}      & c_{13} c_{23} & s^*_{34}   \\[1mm]
        -s_{14}     &  -s_{24}      & -{s'}_{34} & c_{14}c_{24} 
                 \end{array}\right)
\end{eqnarray}
with the definition of $s_{ij}$ and $c_{ij}$ as given in
eq.~(\ref{deltaphases}), and 
\begin{eqnarray}
{s'}_{12}=s_{12}+s_{13} s^*_{23} + s_{14} s^*_{24},\quad
{s'}_{34}=s_{34}+s^*_{13} s_{14} + s^*_{23} s_{24}
\end{eqnarray}
In this approximation, the rotation angles and the phases in ${\sf D}'$ 
can be written as  
\begin{eqnarray}
&& s_{12}=+\frac{m_Z^2\, c_W s_W \left[\,|M_1|\,(|M_2| z^*_1 z_2 
                 + |M_1| z_1 z^*_2) +|\mu|\cos 2\eta\, (|M_2| z_1 z_2 
		 + |M_1| z^*_1 z^*_2)\, \right]}{
		 (|M_2|^2-|M_1|^2)(|M_1|^2-|\mu|^2)}\nonumber\\
&& s_{13}=-\frac{m_Z\, s_W\, c_\eta}{|M_1|^2-|\mu|^2}
           \left(|M_1| z^*_1 + |\mu| z_1\right)\hskip 2cm
   s_{14}=-\frac{m_Z\, s_W\, s_\eta}{|M_1|^2-|\mu|^2}
           \left(|M_1| z^*_1 - |\mu| z_1\right)\nonumber\\
&& s_{23}=+\frac{m_Z\, c_W\, c_\eta}{|M_2|^2-|\mu|^2}
           \left(|M_2| z^*_2 + |\mu| z_2\right) \hskip 2cm
   s_{24}=+\frac{m_Z\, c_W\, s_\eta}{|M_2|^2-|\mu|^2}
           \left(|M_2| z^*_2 - |\mu| z_2\right)\nonumber\\
&& s_{34}=\frac{m_3-|\mu|}{2|\mu|}\tan\eta
         -i \frac{c_\eta\, s_\eta}{m_3-|\mu|}
           \left[X_1 |M_1| \sin(\Phi_1+\Phi_\mu)
                +X_2 |M_2| \sin\Phi_\mu \right]
\end{eqnarray}
where for the sake of notation the parameters 
\begin{eqnarray}
z_1={\rm e}^{-\frac{i}{2}(\Phi_1+\Phi_\mu)}\qquad {\rm and}\qquad 
z_2={\rm e}^{-\frac{i}{2}\Phi_\mu}
\end{eqnarray}
have been introduced. 
On the other hand, the phases $\alpha_i$ in ${\sf
  M}$, 
\begin{eqnarray}
&& \alpha_1 = -\frac{X_1|\mu|}{2 m_1}
               \sin(\Phi_1+\Phi_\mu)\cos 2\eta\nonumber\\
&& \alpha_2 = -\frac{X_2|\mu|}{2 m_2}
               \sin\Phi_\mu\cos 2\eta \nonumber\\
&& \alpha_3 = c^2_\eta\frac{X_1 |M_1| \sin(\Phi_1+\Phi_\mu)
                        +X_2 |M_2| \sin\Phi_\mu}{2 m_3}\nonumber\\
&& \alpha_4 = \frac{\pi}{2}
         - s^2_\eta\frac{X_1 |M_1| \sin(\Phi_1+\Phi_\mu)
                        +X_2 |M_2| \sin\Phi_\mu}{2 m_4}
\end{eqnarray}
are 
expressed in terms of the invariant phases $\Phi_1$ and 
$\Phi_\mu$.
\vskip 6mm

\noindent {\bf Addendum: Charginos}\\[3mm]
The same approximation can be applied to the chargino system. The  
mass matrix \cite{R1}
\begin{eqnarray}
{\cal M}_C = \left(\begin{array}{cc}
     M_2              & \sqrt{2} m_W\, c_{\beta} \\[2mm]
\sqrt{2} m_W\, s_{\beta} & |\mu|\, {\rm e}^{i\Phi_\mu}
             \end{array}\right)
\end{eqnarray}
is diagonalized by two different unitary matrices
$ {U}_R{\cal M}_C {U}^\dagger_L ={\sf diag}\{m^\pm_1,\,
m^\pm_2\}$ 
parameterized in general by two rotation
angles and four phases:
\begin{eqnarray}
{ U}_L = \left(\begin{array}{cc}
        c_{_L}  &  s^*_{_L}  \\
       -s_{_L}  &  c_{_L}
            \end{array}\right)\;\; \mbox{ ~~ and~~  }\;\; 
{ U}_R = {\sf diag}\{\,{\rm e}^{i\gamma_1},\,\, {\rm e}^{i\gamma_2}\}
      \left(\begin{array}{cc}
        c_{_R}  &  s^*_{_R}  \\
       -s_{_R}  &  c_{_R}
            \end{array}\right)
\end{eqnarray}
where $c_{_{L,R}} = \cos\phi_{_{L,R}}$ and 
$s_{_{L,R}}=\sin\phi_{_{L,R}}\, {\rm e}^{i\,\delta_{L,R}}$. The exact
solutions were given in Ref.~\cite{CDGKSZ}. 
In the limit of $M_2^2,\,|\mu|^2\gg m_Z^2$  and $|M_2\pm |\mu||^2\gg
m_Z^2$, the following expressions 
\begin{eqnarray}
&& m^\pm_1= M_2 +X_2\left[\,M_2+|\mu|\,s_{2\beta}\cos\Phi_\mu\right]
   \nonumber\\[2mm]
&& m^\pm_2= |\mu|\,-X_2\left[\,|\mu|+M_2\, s_{2\beta}\cos\Phi_\mu\right]
\end{eqnarray}
are found for the chargino masses and 
\begin{eqnarray}
&&s_{_L}=\frac{\sqrt{2}m_W}{M^2_2-|\mu|^2}\,(M_2\,c_\beta+
   \mu^*\,s_\beta) \hskip 2cm
\gamma_1= +X_2\frac{|\mu|}{M_2}\, s_{2\beta} \sin\Phi_\mu
   \nonumber\\
&&  s_{_R} =\frac{\sqrt{2}m_W}{M^2_2-|\mu|^2}\,(\mu c_\beta+M^*_2\,s_\beta)
\hskip 2.2cm
\gamma_2= -X_2\frac{M_2}{|\mu|}\, s_{2\beta} \sin\Phi_\mu
\end{eqnarray}
for the mixing angles and phases.
 
\subsubsection{The case \boldmath{$M_1=M_2$} in the limit  
            \boldmath{$\tan\beta=1$} \label{degcase}} 

When the two soft--breaking SU(2) and U(1) gaugino masses are equal, 
$|M_1|=M_2=M, \Phi_1=0$, and $\tan\beta$ is unity, the electroweak gauge 
symmetry guarantees the existence of two physical neutral states 
which do not mix with the other states and which have  mass 
eigenvalues identical to the moduli $M$ and $|\mu|$. 
As a result, only one gaugino state and one higgsino state mix with
each other so that   a complete analytic expressions can be derived
for the mass spectrum and the mixing matrix.
For the sake of convenience, the following notation is introduced:
\begin{eqnarray}
&& \lambda = {M}/{m_Z},\quad \nu  = {|\mu|}/{m_Z},\quad
   \Delta  = \left\{(\lambda^2-\nu^2)^2+4(\lambda^2+\nu^2+2\lambda\nu\,
             \cos\Phi_\mu)\right\}^{1/2}\nonumber\\[2mm]
&& \cos\theta = \sqrt{\frac{\Delta-\lambda^2+\nu^2}{2\Delta}}\qquad
   \hskip 2cm
   \sin\theta = \sqrt{\frac{\Delta+\lambda^2-\nu^2}{2\Delta}}\nonumber\\
&& \cos\delta = \frac{2(\nu+\lambda)}{\sqrt{\Delta^2-(\nu^2-\lambda^2)^2}}
                \,\cos\frac{\Phi_\mu}{2}\qquad
   \sin\delta = \frac{2(\nu-\lambda)}{\sqrt{\Delta^2-(\nu^2-\lambda^2)^2}}
                \,\sin\frac{\Phi_\mu}{2}
\end{eqnarray}
With this notation, the neutralino masses $m_i$ are given by 
\begin{eqnarray}
&&  m_1= M\hskip 2.3cm m_2=
  \sqrt{\frac{\lambda^2+\nu^2+2-\Delta}{2}}\, m_Z   \nonumber\\
&& m_4 = |\mu| \hskip 2.3cm m_3 =
  \sqrt{\frac{\lambda^2+\nu^2+2+\Delta}{2}}\, m_Z  
\end{eqnarray}
and the unitary mixing matrix $N= {\sf M}\,  {\sf D}'\, {\sf P}$, 
as defined in eq.~(\ref{eq:approxN}),  is obtained from
the matrix ${\sf D}'$ 
\begin{eqnarray}
{\sf D}' =\left(\begin{array}{cccc}
     1 & 0 & 0 & 0 \\
     0 & \cos\theta & -\sin\theta\, {\rm e}^{i\delta} & 0 \\
     0 & \sin\theta\, {\rm e}^{-i\delta} & \cos\theta & 0 \\
     0 & 0 & 0 & 1
          \end{array}\right)
\end{eqnarray}
and the phase matrix ${\sf M}$ with 
\begin{eqnarray}
&& \alpha_1=0 \hskip 2cm
\alpha_2 = {\sf Arg}\left[1
    -\frac{\nu(\Delta-\nu^2+\lambda^2)}{\lambda(\Delta+\nu^2-\lambda^2)}
          \, {\rm e}^{-2i\delta}\right] \nonumber\\
&& \alpha_4=\frac{\pi}{2} \hskip 1.9cm \alpha_3 = {\sf Arg}\left[1
    -\frac{\lambda(\Delta-\nu^2+\lambda^2)}{\nu(\Delta+\nu^2-\lambda^2)}
          \, {\rm e}^{2i\delta}\,\right] 
\end{eqnarray}
\begin{figure}[htb]
\begin{center}
  \epsfig{file=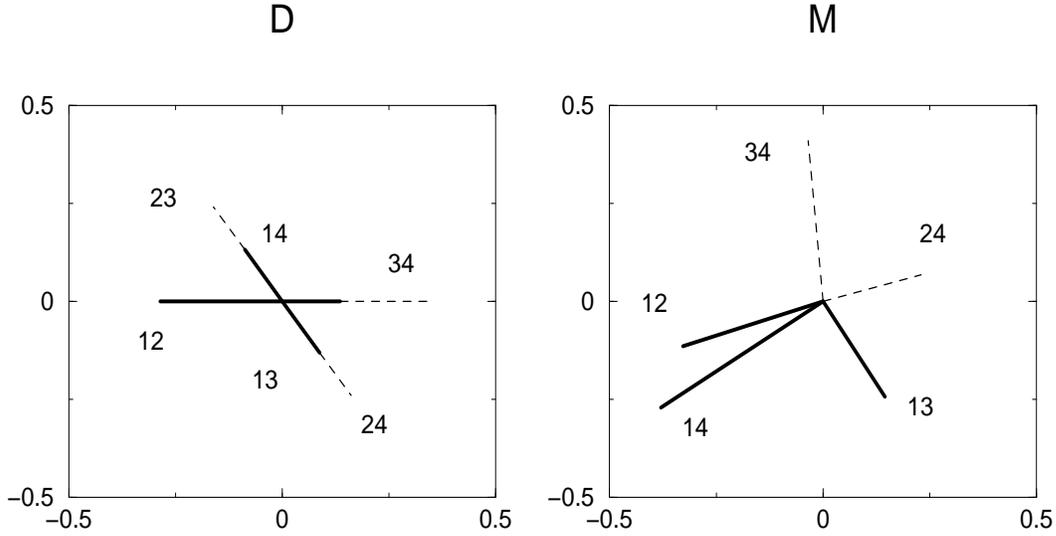,height=14cm,width=7cm,angle=-90} 
  \vspace*{0.4cm}
\caption{\it The $D$--type (left panel) and $M$--type (right panel)
  quadrangles in  the complex
  plane for  the special 
         case of $\tan\beta=1$ and $M_1=M_2=100$ GeV, and $|\mu|=150$ GeV,
         $\Phi_{\mu}=\pi/2$. The quadrangle $M_{23}$  degenerates to a point.} 
\label{quaddeg}
\end{center}
\end{figure}

From the explicit form of the mixing matrix $N$ it is apparent that all
unitarity quadrangles collapse to lines as shown in Fig.\ref{quaddeg}.
However, since the phases $\delta$, and $\alpha_2$ and $\alpha_3$ are
in general non--vanishing, not all lines are parallel to the real or
imaginary axes, a characteristic feature which signals CP--violation.
Only in the CP--conserving case, {\it i.e.} for $\Phi_{\mu}=0$ in this
particular example, the phases $\delta$ vanish (modulo $\pi$) and 
$\alpha_i$ vanish (modulo $\pi/2$) and all
collapsed quadrangles are oriented along the real or the imaginary axis.

\section{Neutralino production in \boldmath{$e^+e^-$} collisions}

\begin{figure}
{\color{blue}
\begin{center}
\begin{picture}(330,100)(10,0)
\Text(5,85)[]{$e^-$}
\ArrowLine(0,75)(25,50)
\ArrowLine(25,50)(0,25)
\Text(5,15)[]{$e^+$}
\Photon(25,50)(65,50){2}{6}
\Text(45,37)[]{\color{red} $Z$}
\Line(65,50)(90,75)
\Photon(65,50)(90,75){2}{6}
\Text(87,87)[]{$\tilde{\chi}^0_i$}
\Line(90,25)(65,50)
\Photon(90,25)(65,50){2}{6}
\Text(87,13)[]{$\tilde{\chi}^0_j$}
\Text(125,85)[]{$e^-$}
\ArrowLine(120,75)(165,75)
\Text(125,15)[]{$e^+$}
\ArrowLine(165,25)(120,25)
\Line(164,75)(164,25)
\Line(166,75)(166,25)
\Text(150,50)[]{\color{red} $\tilde{e}_{L,R}$}
\Line(165,75)(210,75)
\Photon(165,75)(210,75){2}{6}
\Text(207,87)[]{$\tilde{\chi}^0_i$}
\Line(210,25)(165,25)
\Photon(210,25)(165,25){2}{6}
\Text(207,13)[]{$\tilde{\chi}^0_j$}
\Text(245,85)[]{$e^-$}
\ArrowLine(240,75)(285,75)
\Text(245,15)[]{$e^+$}
\ArrowLine(285,25)(240,25)
\Line(284,75)(284,25)
\Line(286,75)(286,25)
\Text(270,50)[]{\color{red} $\tilde{e}_{L,R}$}
\Line(285,75)(330,25)
\Photon(285,75)(330,25){2}{8}
\Text(327,87)[]{$\tilde{\chi}^0_i$}
\Line(330,75)(285,25)
\Photon(330,75)(285,25){2}{8}
\Text(327,13)[]{$\tilde{\chi}^0_j$}
\end{picture}\\
\end{center}
}
\caption{\it Mechanisms contributing to the production of 
  diagonal and non--diagonal neutralino pairs in $e^+e^-$
  annihilation, $e^+e^-\rightarrow \tilde{\chi}^0_i
  \tilde{\chi}^0_j$ $(i,j$=1,2,3,4).}
\label{fig:diagrams}
\end{figure}
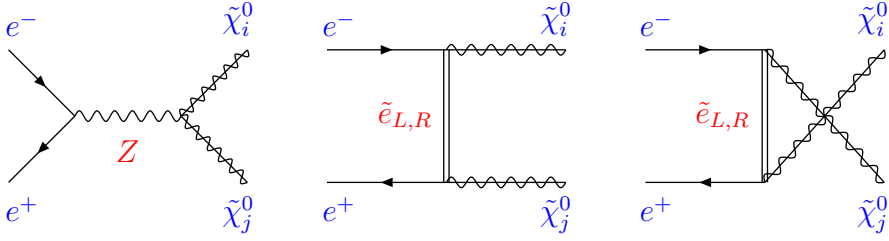

The production processes 
\begin{equation}
e^+e^-\rightarrow \tilde{\chi}^0_i\tilde{\chi}^0_j \qquad (i,j=1,2,3,4) 
\end{equation}
are generated by the five mechanisms shown in Fig.\ref{fig:diagrams}:
$s$-channel $Z$ exchange, and $t$- and $u$-channel $\tilde{e}_{L,R}$
exchanges\footnote{For the reader's convenience, we report
some technical material in chapter 3.1 in parallel to 
Refs.\cite{CDDKZ,CDGKSZ,CSS} so that the presentation becomes
self-contained.}.  The transition matrix element, after an appropriate
Fierz transformation of the $\tilde{e}_{L,R}$ exchange amplitudes,
\begin{eqnarray}
T\left(e^+e^-\rightarrow\tilde{\chi}^0_i\tilde{\chi}^0_j\right)
 = \frac{e^2}{s}\, Q_{\alpha\beta}
   \left[\bar{v}(e^+)  \gamma_\mu P_\alpha  u(e^-)\right]
   \left[\bar{u}(\tilde{\chi}^0_i) \gamma^\mu P_\beta 
               v(\tilde{\chi}^0_j)\right]
\label{eq:neutralino production amplitude}
\end{eqnarray}
can be expressed in terms of four generalized bilinear charges
$Q_{\alpha\beta}$. They correspond to independent helicity 
amplitudes \cite{SZ} 
which describe the neutralino production processes for polarized
electrons/positrons (the lepton mass  neglected).
They  are defined by the lepton and
neutralino currents and the propagators of the exchanged (s)particles
as  follows:
\begin{eqnarray}
&& Q_{LL}=+\frac{D_Z}{s_W^2c_W^2}\,
           (s_W^2-{\textstyle{\frac{1}{2}}}\,) {\cal Z}_{ij}
            -D_{uL}g_{Lij}\qquad
 Q_{RL}=+\frac{D_Z}{c_W^2}\, 
            {\cal Z}_{ij}
            +D_{tR}g_{Rij}\nonumber\\ 
&& Q_{LR}=-\frac{D_Z}{s_W^2c_W^2}\, 
           (s_W^2 -{\textstyle{\frac{1}{2}}}\,){\cal Z}^*_{ij}
            +D_{tL}g^*_{Lij}\qquad 
 Q_{RR}=-\frac{D_Z}{c_W^2}{\cal Z}^*_{ij}
            -D_{uR}g^*_{Rij}
\end{eqnarray}
The first index in $Q_{\alpha\beta}$ refers to the chirality of the
$e^\pm$ current, the second index to the chirality of the
$\tilde{\chi}^0$ current. The first term in each bilinear charge is
generated by $Z$--exchange and the second term by selectron
exchange; $D_Z$, $D_{tL,R}$ and $D_{uL,R}$ denote the $s$--channel Z
propagator and the $t$-- and $u$--channel left/right--type selectron
propagators
\begin{eqnarray}
&& D_Z=\frac{s}{s-m^2_Z+im_Z\Gamma_Z}\nonumber\\
&& D_{tL,R}=\frac{s}{t-m^2_{\tilde{e}_{L,R}}}  \qquad {\rm and }
\qquad t\rightarrow u 
\end{eqnarray}
with $s=(p_{e^-}+p_{e^+})^2$, $t=(p_{e^-}-p_{\tilde{\chi}^0_i})^2$ and
$u=(p_{e^-}-p_{\tilde{\chi}^0_j})^2$. The matrices ${\cal Z}_{ij}$,
$g_{Lij}$ and $g_{Rij}$ are products of the neutralino diagonalization
matrix elements $N_{ij}$
\begin{eqnarray}
&& {\cal Z}_{ij}=
                 (N_{i3}N^*_{j3}-N_{i4}N^*_{j4})/2\nonumber\\
&& g_{Lij}=(N_{i2}c_W+N_{i1}s_W)(N^*_{j2}c_W+N^*_{j1}s_W)/
                4 s_W^2c_W^2\nonumber\\
&& g_{Rij}=N_{i1}N^*_{j1}/c_W^2
\end{eqnarray}
They satisfy the hermiticity relations reflecting the CP relations 
\begin{eqnarray}
{\cal Z}_{ij}={\cal Z}^*_{ji}\qquad
g_{Lij}=g^*_{Lji}\qquad
g_{Rij}=g^*_{Rji}
\end{eqnarray}
so that, if the $Z$--boson width $\Gamma_Z$ is neglected in the 
$Z$--boson propagator $D_Z$, the bilinear charges $Q_{\alpha\beta}$ 
also satisfy  similar relations
with $t$ and $u$ interchanged in the propagators. These relations are
very useful in classifying CP--even and CP--odd observables
in the following sections.

\subsection{Production cross sections}

Since the gaugino and higgsino interactions depend on the chirality of
the states, polarized electron and positron beams are useful
tools to diagnose  the wave-functions of the neutralinos.  The
electron and positron polarization vectors are defined in the reference
frame in which  the electron--momentum direction
defines  the $z$--axis and
the electron transverse polarization--vector the $x$--axis.  The
azimuthal angle of the transverse polarization--vector of the positron
with respect to the $x$--axis is called $\eta$. The
polarized differential cross section for the $\tilde{\chi}^0_i
\tilde{\chi}^0_j$ production 
is given in terms of the electron $P$=$(P_T,0,P_L)$
and positron $\bar{P}$=$(\bar{P}_T \cos\eta,\bar{P}_T\sin\eta,
-\bar{P}_L)$ polarization vectors  
 by
\begin{eqnarray}
&& \frac{{\rm d}\sigma}{{\rm d}\Omega}\{ij\}
  =\frac{\alpha^2}{16\, s}\, \lambda^{1/2} \bigg[
     (1-P_L\bar{P}_L)\,\Sigma_U+(P_L-\bar{P}_L)\,\Sigma_L
     \nonumber\\
&& { }\hskip 3cm 
  +P_T\bar{P}_T\cos(2\Phi-\eta)\,\Sigma_T
  +P_T\bar{P}_T\sin(2\Phi-\eta)\,\Sigma_N\bigg]\ \label{eq:diffx}
\end{eqnarray}
with the coefficients $\Sigma_U$, $\Sigma_L$, $\Sigma_T$
and $\Sigma_N$ depending 
only on the polar angle $\Theta$ of the produced neutralinos, 
but not on the azimuthal 
angle $\Phi$ any more; $\lambda=[1-(\mu_i+\mu_j)^2][1-(\mu_i-\mu_j)^2]$ 
is the two--body phase space function with $\mu_i=m_{\tilde\chi_i^0}/\sqrt{s}$.
The coefficients $\Sigma_U$, $\Sigma_L$, $\Sigma_T$ and
$\Sigma_N$ are written in terms of the quartic charges

\begin{eqnarray}
&& \Sigma_{U}=4\left\{\left[1-(\mu^2_i - \mu^2_j)^2
                   +\lambda\cos^2\Theta\right]Q_1
                   +4\mu_i\mu_j Q_2+2\lambda^{1/2} Q_3\cos\Theta\right\}
                  \nonumber\\
&& \Sigma_{L}=4\left\{\left[1-(\mu^2_i - \mu^2_j)^2
                   +\lambda\cos^2\Theta\right]Q'_1
                   +4\mu_i\mu_j Q'_2+2\lambda^{1/2} Q'_3\cos\Theta\right\}
                  \nonumber\\
&& \Sigma_{T}=4\lambda \, Q_5 \sin^2\Theta \nonumber \\ 
&& \Sigma_{N}=4\lambda \, Q'_6 \sin^2\Theta
\label{eq:initial}
\end{eqnarray}
Expressed in terms of 
bilinear charges, the quartic charges are collected in Table~1, including the
transformation properties under P~and~CP. 
\begin{table*}[\hbt]
\caption[{\bf Table 1:}]{\label{tab:quartic} 
{\it The independent quartic charges of the neutralino system.}}
\begin{center}
\begin{tabular}{|c|c|l|}\hline
 &  &  \\[-4mm]
${\rm P}$ & ${\rm CP}$ & { }\hskip 2cm Quartic charges \\\hline \hline
 &  &  \\[-3mm]
 even    &  even     & $Q_1 =\frac{1}{4}\left[|Q_{RR}|^2+|Q_{LL}|^2
                       +|Q_{RL}|^2+|Q_{LR}|^2\right]$ \\[2mm]
         &           & $Q_2 = \frac{1}{2}\real\left[Q_{RR}Q^*_{RL}
                       +Q_{LL}Q^*_{LR}\right]$ \\[2mm]
         &           & $Q_3 = \frac{1}{4}\left[|Q_{RR}|^2+|Q_{LL}|^2
                       -|Q_{RL}|^2-|Q_{LR}|^2\right]$ \\[2mm]
         &           & $Q_5=\frac{1}{2}\real \left[Q_{LR}Q^*_{RR}
                       +Q_{LL}Q^*_{RL}\right]$ \\
 & & \\[-3mm]
\cline{2-3} 
 & & \\[-3mm]
         &  odd      & $Q_4=\frac{1}{2}\imag\left[Q_{RR}Q^*_{RL}
                       +Q_{LL}Q^*_{LR}\right]$\\[2mm] \hline \hline
 & & \\[-3mm]
 odd     &  even     & $Q'_1=\frac{1}{4}\left[|Q_{RR}|^2+|Q_{RL}|^2
                        -|Q_{LR}|^2-|Q_{LL}|^2\right]$\\[2mm]
         &           & $Q'_2=\frac{1}{2}\real\left[Q_{RR}Q^*_{RL}
                        -Q_{LL}Q^*_{LR}\right]$ \\[2mm]
         &           & $Q'_3=\frac{1}{4}\left[|Q_{RR}|^2+|Q_{LR}|^2
                        -|Q_{RL}|^2-|Q_{LL}|^2\right]$\\
 & & \\[-3mm]
\cline{2-3} 
 & & \\[-3mm]
         &  odd      & $Q'_6=\frac{1}{2}\imag\left[Q_{RR}Q^*_{LR}
                       -Q_{LL}Q^*_{RL}\right]$\\[2mm] 
\hline
\end{tabular}
\end{center}
\end{table*}

The quartic charges $Q_4\{ij\}$ and  $Q'_6\{ij\}$, which are 
non--vanishing only for
$i\neq j$ and for CP--violating theories, can be expressed in terms
of the elements of the 
mixing matrix $N$. Taking the $Z$-boson propagator real by
neglecting the width in the limit of high energies, 
the quartic charge $Q'_6\{ij\}$ is given by 
\begin{eqnarray}
&& Q'_6\{ij\}= \frac{D_Z}{2s_W^2 c_W^2}
          \bigg[\,s_W^2 (D_{tL}-D_{uL})\,\imag({\cal Z}_{ij} g^*_{Lij})
          -\left(s_W^2-\textstyle{\frac{1}{2}}\right)(D_{tR}-D_{uR})
	  \,\imag({\cal Z}_{ij}g^*_{Rij})\,\bigg]
           \nonumber\\
      && \hskip 3.5cm+ \frac{1}{2}\left(D_{tL}D_{uR}-D_{tR}D_{uL}\right)
           \imag(\,g_{Lij}g^*_{Rij})
\end{eqnarray}
The combinations of the couplings, $\imag({\cal Z}_{ij} g^*_{Lij})$,
$\imag({\cal Z}_{ij}g^*_{Rij})$ and $\imag(g_{Lij}g^*_{Rij})$, are 
functions of the plaquettes:
\begin{eqnarray}
&& \imag({\cal Z}_{ij}g^*_{Rij})
   = \frac{1}{2c_W^2}\left[\,\imag(N_{i3}N^*_{j3}N^*_{i1}N_{j1})
                 -\imag(N_{i4}N^*_{j4}N^*_{i1}N_{j1})\,\right]\nonumber\\
&& \imag({\cal Z}_{ij}g^*_{Lij})
   = \frac{1}{8s_W^2c_W^2}\left[\,
                  \imag({N}_{i3}{N}^*_{j3}{N'}^*_{i2}{N'}_{j2})
                 -\imag({N}_{i4}{N}^*_{j4}{N'}^*_{i2}{N'}_{j2})\,\right]
                  \nonumber\\
&& \imag(g_{Lij}g^*_{Rij}) 
   = \frac{1}{4 s^2_W c^4_W}\,\imag({N'}_{i2} {N'^*}_{j2} N^*_{i1} N_{j1})
\label{eq:normal1}
\end{eqnarray}
where $N'_{i1}=c_W N_{i1}-s_W N_{i2}$ and $N'_{i2}=s_W N_{i1}+c_W
N_{i2}$.  The quartic charge $Q_4\{ij\}$ will be discussed in section
\ref{sec:pol}. 

The expression (\ref{eq:normal1}) reveals the following
features: (i) The charge $Q'_6\{ij\}$ vanishes for $i=j$. (ii) 
Non--zero values of 
$\imag({\cal Z}_{ij}g^*_{Rij})$ and $\imag({\cal Z}_{ij} g^*_{Lij})$
require the existence of non-vanishing gaugino and higgsino components
in $\tilde{\chi}^0_i$ and $\tilde{\chi}^0_j$; moreover, the
$\tilde{H}^0_1$ and $\tilde{H}^0_2$ 
higgsino components have to be different in magnitude, which in turn requires
$\tan\beta\neq 1$. (iii) For the transverse beam polarization and
$i\neq j$, the angular distribution
(\ref{eq:diffx}) is forward--backward asymmetric, because the
angular dependence of $\Sigma_N$ is determined by the
forward--backward asymmetric factors, $D_{tL,R}-D_{uL,R}$ and
$D_{tL}D_{uR}-D_{tR}D_{uL}$.

If the neutralino production angle could be measured unambiguously on an 
event--by--event basis, the quartic charges could be extracted 
directly from the angular dependence of the cross section at a 
fixed c.m. energy. However, since the lightest neutralino escapes undetected
and the heavier neutralinos decay into the invisible lightest 
neutralinos as well as SM fermion pairs, the production angle 
cannot be determined unambiguously for non--asymptotic energies. 
However, as a counting experiment, the integrated
polarization--dependent total cross sections can be determined 
unambiguously:
\begin{eqnarray}
&& \sigma_{R}={\cal S}_{ij}\, 
               \int{\rm d}\Omega\,\,\frac{{\rm d}\sigma}{{\rm d}\Omega}
               \left[P_L=-\bar{P}_L=+1\right] \nonumber\\
&& \sigma_{L}={\cal S}_{ij}\, 
               \int{\rm d}\Omega\,\,\frac{{\rm d}\sigma}{{\rm d}\Omega}
               \left[P_L=-\bar{P}_L=-1\right] 
\label{eq:xsections}
\end{eqnarray}
where ${\cal S}_{ij}$ is a statistical factor: 1 for $i\neq j$ and
$1/2$ for $i = j$. 
Twenty independent physical observables can be constructed at a 
given c.m. energy through neutralino--pair production with polarized 
electron and positron beams; two for each mode $\{ij\}$.
The generalization of eq.~(\ref{eq:xsections}) for partially
polarized beams is straightforward.

\begin{center}
\begin{figure}[htb]
\vspace*{0.0cm}
\hspace*{3.0cm}
 \epsfxsize=10cm \epsfbox{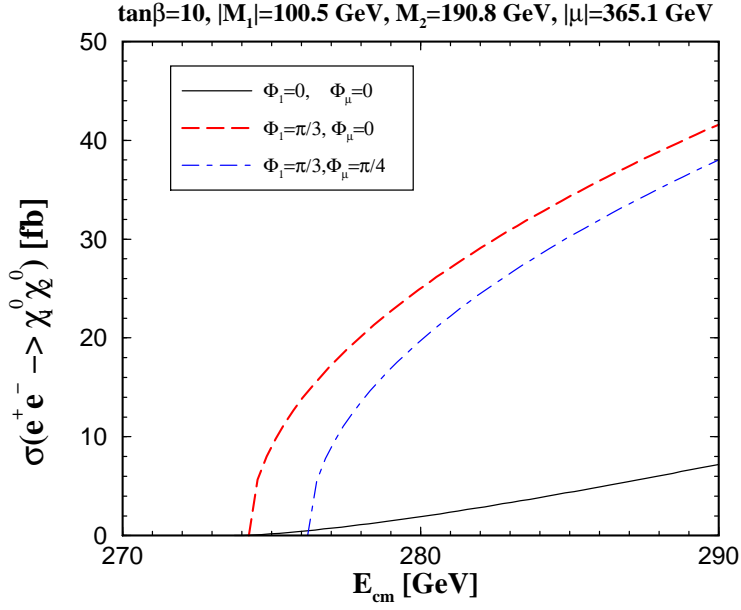}
\caption{\it The threshold behavior of the neutralino  production
  cross--section $\sigma\{12\}$; the shift of the energy threshold is
  due to the dependence of the neutralino masses on the phases.}
\label{fig:th}
\end{figure}
\end{center}
%

\subsection{Threshold behavior of neutralino production}

Near the threshold of each non--diagonal neutralino pair, the 
total cross section $\sigma\{ij\}$ ($i\neq j$) is approximately given
by  
\begin{eqnarray}
&&\sigma\{ij\}\approx \frac{\pi\alpha^2\,\lambda^{1/2}\,}{(m_i+m_j)^2}
          \,\bigg\{\, \frac{4m_im_j}{(m_i+m_j)^2}\, |\imag\, G^{(0)}_R|^2\, 
	   \nonumber \\[2mm]
&& \quad + \lambda\, \bigg[ \,
\frac{2 m_i m_j}{(m_i+m_j)^2}\,\imag\, G^{(0)}_R\imag\, G^{(1)}_R
   +\bigg(\frac{(m_i+m_j)^2}{4m_i m_j}
               -\frac{1}{3}\bigg)\, |G^{(0)}_R|^2\, \nonumber \\[2mm]
&& \qquad +\frac{m_i m_j}{3(m_i+m_j)^2} F_0^4\, |\real\; g_{Rij}|^2
-2\, |\imag\, G^{(0)}_R|^2
+\frac{1}{3} F_0^2 \, \real\, (G^{(0)}_R\, g^*_{Rij})\bigg]\nonumber\\[2mm]
&& \quad
 +\left[G^{(0,1)}_R\!\rightarrow G^{(0,1)}_L, 
        D_0\! \rightarrow D_0(s_W^2-1/2)/s_W^2, 
        g_{Rij}\!\rightarrow\! -g_{Lij}, 
        m_{\tilde{e}_R}\rightarrow m_{\tilde{e}_L}\right]\bigg\}  
\label{eq:thres}
\end{eqnarray}
where 
\begin{eqnarray}
&& G^{(0,1)}_R=\frac{1}{c_W^2}\, D_{0,1} {\cal Z}_{ij}
  - F_{0,1}\, g_{Rij}
\end{eqnarray}
with the kinematical functions
\begin{eqnarray}
&& D_0=(m_i+m_j)^2/((m_i+m_j)^2-m_Z^2)\nonumber \\[1mm]
&& D_1=-m_Z^2 (m_i+m_j)^4/m_i m_j ((m_i+m_j)^2-m_Z^2)^2\nonumber\\[1mm]
&& F_0=(m_i+m_j)^2/(m^2_{\tilde{e}_R}+m_i m_j)\nonumber \\[1mm]
&& F_1=(m_i+m_j)^4\, (2m^2_{\tilde{e}_R}-m_i^2-m_j^2)/ 
2m_i m_j (m^2_{\tilde{e}_R}+m_i m_j)^2 +F_0^3/3
\end{eqnarray}
In the CP--invariant theory, the imaginary parts of the couplings
${\cal Z}_{ij}$, $g_{Lij}$ and $g_{Rij}$ can only be generated by 
Majorana phases $\alpha_i=0$ and $\alpha_j=\pi/2$ or vice versa. 
Therefore the S--wave excitation
giving rise to a steep rise $\sim \lambda^{1/2}$ of the cross section
for the nondiagonal pairs\footnote{For diagonal pairs the couplings 
${\cal Z}_{ii}$, $g_{Lii}$ and $g_{Rii}$ are real.} near threshold,
signals opposite CP--parities of the produced neutralinos \cite{R2}. 
Obviously not all nondiagonal pairs of neutralinos can be produced 
in S--wave in the CP--invariant theory at the same time; if
the $\{ij\}$ and $\{ik\}$ pairs have negative CP--parities, the
pair $\{jk\}$ have positive CP--parity and will be excited 
in a P--wave characterized  by the
slow rise $\sim\lambda^{3/2}$ of the cross section.

It is important to realize that CP--violation may allow S--wave excitations
in all non--diagonal pairs. In particular, observing the   $\{ij\}$,
$\{ik\}$ and $\{jk\}$ pairs to be excited {\it all} in S--wave states would 
therefore signal CP--violation.  
In Fig.\ref{fig:th} the impact of non--zero CP phases $\Phi_1$ and
$\Phi_\mu$ on the threshold behavior of $\sigma\{12\}$ is shown. For
vanishing phases the $\tilde{\chi}^0_1$ and $\tilde{\chi}^0_2$ fields
have the
same CP--parities and thus the production cross section rises as
$\lambda^{3/2}$.  Evidently the CP--violating phases have a strong
impact on the energy dependence of the cross section, as anticipated
in eq.~(\ref{eq:thres}).  Thus, the steep rise of cross sections for
non--diagonal pairs can be interpreted as  a first direct
signature  of the
presence of CP--violation in the neutralino sector.

\subsection{Neutralino polarization vector \label{sec:pol}}

If the initial beams are not polarized, the chiral structure
of the neutralinos could be inferred from the polarization of the 
$\tilde{\chi}^0_i\tilde{\chi}^0_j$ pairs produced in $e^+e^-$ 
annihilation.

The polarization vector $\vec{\cal P}=({\cal P}_L, {\cal P}_T, {\cal
P}_N) $  is defined in the rest frame of the particle
$\tilde{\chi}^0_i$, with components parallel to
the $\tilde{\chi}^0_i$ flight direction in the c.m. frame, in the
production plane, and normal to the production plane, respectively.
They are expressed in terms of the quartic
charges as follows
{\small
\begin{eqnarray}
&& {\cal P}_L =4\left\{ 2(1-\mu^2_i-\mu^2_j)\,\cos\Theta\,Q'_1
              +4\mu_i\mu_j\,\cos\Theta\, Q'_2
              +\lambda^{1/2}[1+\cos^2\Theta-(\mu^2_i-\mu^2_j)]\, Q'_3 
              \right\} / \Sigma_{U} \nonumber \\
&& {\cal P}_T =-8\sin\Theta\left\{[(1-\mu^2_i+\mu^2_j)\,Q'_1
              +\lambda^{1/2}\, Q'_3\cos\Theta]\mu_i
              +(1+\mu^2_i-\mu^2_j)\mu_j\,Q'_2\right\}/\Sigma_{U}\nonumber\\ 
&& {\cal P}_N =8\lambda^{1/2}\mu_j\,\sin\Theta\, Q_4/\Sigma_{U}
\end{eqnarray}}
with the normalization $\Sigma_{U}$ as defined in
eq.~(\ref{eq:initial}).  

The normal component ${\cal P}_N$ can only be
generated by complex production amplitudes. Neglecting the $Z$--boson
width, the normal $\tilde{\chi}^0_i$ polarization in
$e^+e^-\rightarrow \tilde{\chi}^0_i \tilde{\chi}^0_i$ is zero since
the $Z\tilde{\chi}_i \tilde{\chi}_i $ vertices and the
selectron--exchange amplitudes are real even for non-zero phases in
the neutralino mass matrix. Only for nondiagonal $\tilde{\chi}^0_i
\tilde{\chi}^0_j$ pairs with $i\neq j$ the amplitudes can be complex
giving rise to a non--zero CP--violating normal neutralino
polarization ${\cal P}_N$ determined by the quartic charge
\begin{eqnarray}
\lefteqn{Q_4\{ij\}=\frac{D^2_Z}{2s_W^4 c_W^4}\,
      \left(s^2_W-\textstyle{\frac{1}{4}}\right)\,
      \imag\left({\cal Z}^2_{ij}\right)+D_{uR}D_{tR} \imag(g^2_{Rij})
      -D_{uL}D_{tL}\imag(g^2_{Lij})} \hskip 2cm\nonumber\\
      && \hskip -1.5cm +\frac{D_Z}{s^2_W c^2_W} 
      \left[s^2_W (D_{tR}+D_{uR})\imag({\cal Z}_{ij} g_{Rij})
           +\left(s^2_W-\textstyle{\frac{1}{2}}\right)(D_{tL}+D_{uL})
            \imag({\cal Z}_{ij}g_{Lij})\right] 
\end{eqnarray}
Since $s^2_W=\sin^2\theta_W$ is close to $\frac{1}{4}$, the $Z$--exchange
contribution to the quartic charge $Q_4\{ij\}$ is suppressed.
Nevertheless, unless selectrons are very heavy and CP is conserved,
the normal polarization of the neutralino will provide a crucial
diagnostic probe of CP--violation in the neutralino sector.
Furthermore, the normal polarization signals the existence of
non--trivial $\alpha$-type CP phases so that it can be non--zero even
if all the $\delta$-type CP phases vanish, {\it i.e.} if all the
quadrangles of the neutralino mixing matrix collapse to lines 
with at least one line off the
real and imaginary axes.

\section{Extracting the fundamental SUSY parameters }

The fundamental SUSY parameters
can be extracted from the gaugino-higgsino  sector at an $e^+e^-$
linear collider with an energy  $\sqrt{s}=500$ to 800 GeV. 

The numerical analyses presented below have been worked out for one  
parameter point\footnote{This point corresponds to one of the mSUGRA points
chosen as reference points at the Snowmass Workshop 2001 after combining "Les
Points d'Aix" with part of the CERN points \cite{RR2}.}
in the CP--invariant case and two related
parameter points in the CP--noninvariant case:  
{\small
\begin{eqnarray}
&& {\sf RP1}\,\,:\,
  (\tan\beta, |M_1|, M_2,|\mu|,\, \Phi_1, \Phi_\mu)
             =(10,\,100.5\,{\rm GeV},\, 190.8\, {\rm GeV},\,
                365.1\, {\rm GeV},\,0,\,\, 0)
  \nonumber\\[2mm]
&& {\sf RP1'}\,:\,
  (\tan\beta, |M_1|, M_2,|\mu|,\,\Phi_1, \Phi_\mu)
             =(10,\, 100.5\,{\rm GeV},\, 190.8\, {\rm GeV},\,
                365.1\, {\rm GeV},\, \frac{\pi}{3},\,0)
  \nonumber\\
&& {\sf RP1''}:\,
  (\tan\beta, |M_1|, M_2, |\mu|,\, \Phi_1, \Phi_\mu)
             =(10,\,100.5\,{\rm GeV},\, 190.8\, {\rm GeV},
                 365.1\, {\rm GeV},\,\frac{\pi}{3},\,\frac{\pi}{4})
\label{eq:parameter}
\end{eqnarray}
}
\noindent
\hskip -0.1cm 
The induced neutralino $\tilde{\chi}^0_i$  masses read as follows
\begin{eqnarray}
\begin{array}{ll}
   m_{\tilde{\chi}^0_1}=97.6/98.2/99.1\,{\rm GeV}\,\hskip 1cm &
   m_{\tilde{\chi}^0_2}=176.2/176.0/177.0\,{\rm GeV} \\[1mm]
   m_{\tilde{\chi}^0_3}=371.4/371.7/372.0\,{\rm GeV}\,  \hskip 1cm &
   m_{\tilde{\chi}^0_4}=388.9/388.5/387.5\,{\rm GeV} \\[1mm]
\end{array}
\label{eq:chimasses}
\end{eqnarray}
for the three points ${\sf RP1/1'/1''}$, respectively, 
and the selectron $\tilde{e}_{L,R}$ masses are taken as 
\begin{eqnarray}
\begin{array}{ll}
   m_{\tilde{e}_L}=208.7\,{\rm GeV}\,    \hskip 1cm &
   m_{\tilde{e}_R}=144.1\,{\rm GeV}
\end{array}
\label{eq:semasses}
\end{eqnarray}
for all three points.  Although the first point ${\sf RP1}$
has been defined for an intermediate $\tan\beta$ solution of universal gaugino
and scalar masses at the GUT scale, we decouple our strictly
low--energy phenomenological analysis from the origin and  use
the parameters in eq.~(\ref{eq:parameter}) as just--so input for the
neutralino spectra and couplings.
For the ${\sf RP1'}$ point, only the phase of $M_1$ is non-zero while 
the chargino sector is  CP--conserving, as
suggested  by the EDM constraints \cite{CPmad}. Finally, in ${\sf RP1''}$
both $M_1$ and $\mu$ have large phases. This point
is taken just for illustrative purpose. 

The masses of the selectrons are assumed to be known from
threshold scans in pair production \cite{martyn} or, if $\tilde{e}_L$
is not accessible in direct production but only $\tilde{\nu}$, 
by means  of the SUSY relation $
m^2_{\tilde{e}_L}-m^2_{\tilde{\nu}}=-m_W^2 c_{2\beta} $ fulfilled
exactly at tree level. Complementary tests can be made by studying
forward--backward asymmetries of the decay leptons of neutralinos
\cite{Gudi}.

\begin{figure}[tb]
\begin{center}
 \epsfig{file=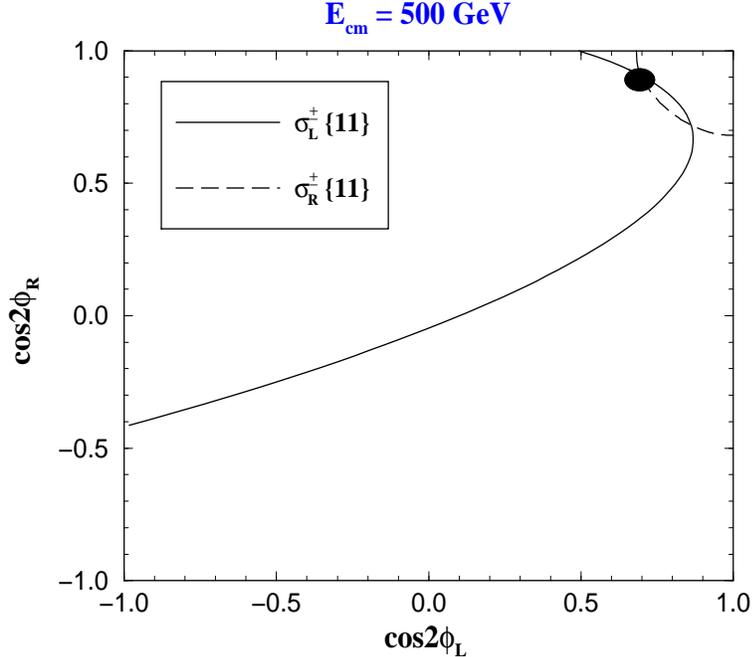,height=9cm,width=10cm}
\caption{\it Contours of the chargino production cross--sections
           $\sigma^\pm_L\{11\} = 341.1$ fb and $\sigma^\pm_R\{11\}=0.53$ fb 
	   for the light chargino mass $m_{\tilde{\chi}^\pm_1}=175.6$ GeV 
	   and the sneutrino mass $m_{\tilde{\nu}}=192.8$ GeV 
           (the set ${\sf RP1}$) in the plane 
	   of $\{\cos 2\phi_L,\, \cos 2\phi_R\}$ at the $e^+e^-$ c.m. 
	   energy of 500 GeV\,; the two crossing points
	   in the upper right corner are $\{0.699,\, 0.906\}$ and
	  $\{0.862,\, 0.720\}$, respectively.}
\label{fig:light}
\end{center}
\end{figure}

\subsection{Light chargino and neutralino system}

%
At the beginning of future $e^+e^-$ linear--collider operations, the
energy may only be sufficient to reach the threshold of the light
chargino pair $\tilde{\chi}^+_1\tilde{\chi}^-_1$ and of the neutralino
pair $\tilde{\chi}^0_1\tilde{\chi}^0_2$.\footnote{The lightest
  neutralino--pair production is difficult to reconstruct experimentally
  but photon tagging in the reaction $e^+e^-
  \rightarrow \gamma \tilde{\chi}^0_1\tilde{\chi}^0_1$ \cite{photon}
  provides a possible method.}
From the analysis of this restricted system,  the
entire structure of the gaugino/higgsino sector can be unraveled
in CP--invariant theories on which we focus first for the sake of simplicity.  
As shown in Ref.~\cite{CDGKSZ},  the chargino sector 
can be reconstructed up to at most a two--fold discrete
ambiguity.  On the other hand, if the
analysis of the chargino and the neutralino systems is combined, 
ten physical observables can be measured: three masses and seven polarized
cross sections, among which two masses and four cross sections are
accessible in the neutralino system.

\begin{figure}[tb]
\begin{center}
 \epsfig{file=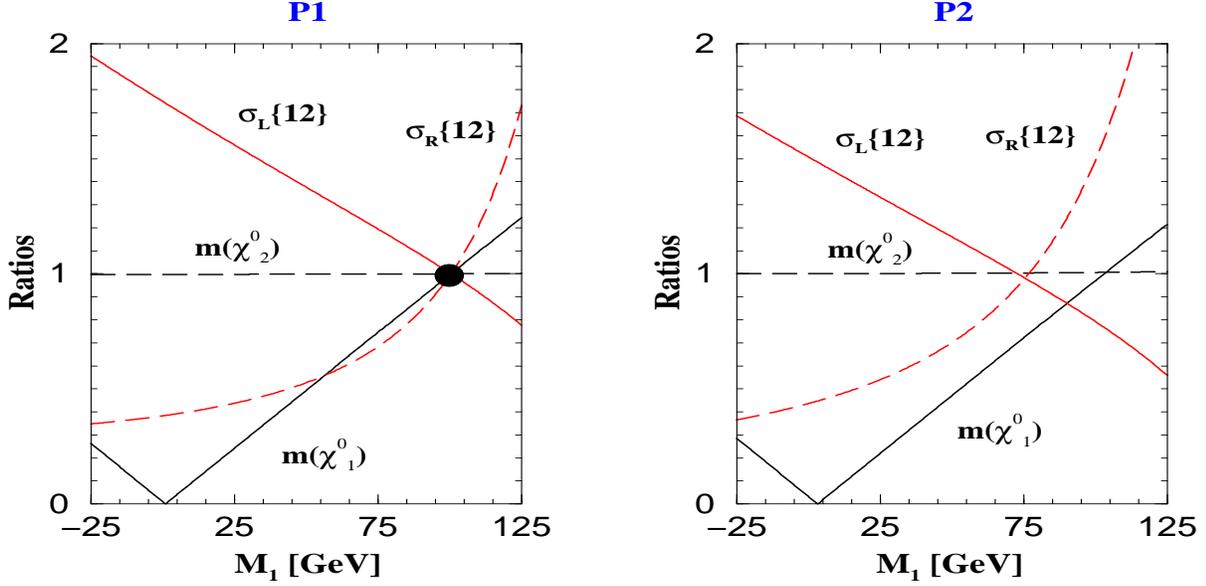,height=8cm,width=16cm}
\caption{\it Ratios of $m_{\tilde{\chi}^0_1}$,
$m_{\tilde{\chi}^0_2}$, $\sigma_L\{12\}$ and $\sigma_R\{12\}$ with
respect to their measured values plotted as functions of $M_1$ for two
possible solutions ${\sf P1}$ (left) and ${\sf P2}$ (right) derived
from the chargino sector.  The left panel gives a unique value
$M_1=100.5$ GeV for the U(1) gaugino mass resolving the 
${\sf P1}$--${\sf P2}$ ambiguity.}
\label{fig:nmass}
\end{center}
\end{figure}

By analyzing the $\{11\}$ mode in $\sigma^\pm_L\{11\}$ and
$\sigma^\pm_R\{11\}$, the chargino mixing angles $\cos2\phi_L$ and
$\cos2\phi_R$ can be determined up to at most a four--fold ambiguity
if the sneutrino mass is known and the SUSY Yukawa coupling is identified
with the gauge coupling.  The ambiguity can be resolved \cite{CDGKSZ}
by measuring\footnote{The measurement of the transverse cross section 
involves the azimuthal production angle $\Phi$ of the charginos. 
At very high energies their angle
coincides with the azimuthal angle of the chargino decay products. With
decreasing energy, however, the angles differ and the measurement of 
the transverse cross section is diluted.}
the transverse cross--section $\sigma^\pm_T\{11\}$.  
On the other hand, 
initial beam polarization in the process $e^+e^-\rightarrow
\tilde{\chi}^0_1\, \tilde{\chi}^0_2$ allows us to measure the two
independent additional observables $\sigma_R\{12\}$ and $\sigma_L\{12\}$
in the neutralino system. Moreover,
the light neutralino masses can be measured with
high precision.

For illustration, we  assume that at the c.m. energy
$E_{cm}=500$ GeV the light chargino mass and 
the polarized cross sections of the light chargino pair are measured with 
good precision to be
$m_{\tilde{\chi}^\pm_1}= 175.6\, {\rm GeV}$ and  
$\sigma^\pm_{L/R}\{11\} = 341.1\, {\rm fb}/0.53\, {\rm fb}$
and the sneutrino mass $m_{\tilde{\nu}}=192.8$ GeV, corresponding to 
${\sf RP1}$. 

The two ellipses in Fig.\ref{fig:light} for the measured 
polarized cross sections $\sigma^\pm_{L,R}\{11\}$, as functions of  
$\cos 2\phi_L$ and $\cos 2\phi_R$, cross at two points:
\begin{eqnarray}
\{\cos 2\phi_{_L},\, \cos 2\phi_{_R}\} = \{0.699,\, 0.906\}\ \ {\rm and}\ \
                                  \{0.862,\, 0.720\}
\label{eq:cosines}
\end{eqnarray}
Following the analysis  described in Ref.~\cite{CDGKSZ},
the cosines of the two mixing angles in eq.~(\ref{eq:cosines}) and the 
light chargino mass $m_{\tilde{\chi}^\pm_1}=175.6$ GeV are sufficient to 
solve for the fundamental parameters
$\{\tan\beta,\, M_2,\, \mu\}$: 
\begin{eqnarray}
\begin{array}{llcl}
{\sf P1}\, : & \{0.699,\, 0.906\} & \Rightarrow 
   & \{\tan\beta=\ \ 10;\, M_2=190.8\,{\rm GeV},\,\mu=365.1\,{\rm GeV}\}\\[2mm]
{\sf P2}\, : & \{0.862,\, 0.720\} & \Rightarrow 
   & \{\tan\beta=0.35;\, M_2=197.9\, {\rm GeV},\, \mu=387.7\, {\rm GeV}\} 
\end{array} \label{twosol}
\end{eqnarray}
The ambiguity can be resolved in several ways: internally within the
chargino sector by measuring the transverse cross--section
$\sigma^\pm_T\{11\}$; externally by confronting the ensuing Higgs
boson mass $m_{h^0}$ with the experimental value. However, the
ambiguity can also be resolved by analyzing the  $\tilde{\chi}^0_1\,
\tilde{\chi}^0_2$  system for left and right polarized beams;    
at the same time the U(1) gaugino mass
parameter can be determined unambiguously.

We assume the measured light neutralino and selectron masses to be those
in eqs.~(\ref{eq:chimasses}) and (\ref{eq:semasses}) and
the measured polarized cross sections $\sigma_{L,R}\{12\}$ to be
233.4 fb /22.1 fb, respectively,  as predicted in ${\sf RP1}$. The expected
values  of $m_{\tilde{\chi}^0_1}$, 
$m_{\tilde{\chi}^0_2}$, $\sigma_L\{12\}$ and $\sigma_R\{12\}$ for the
two possible solutions of eq.~(\ref{twosol})  can be
calculated as functions 
of $M_1$ and compared 
with measured values. In Fig.~\ref{fig:nmass}  the ratios 
of the theoretically predicted values $m_{\tilde{\chi}^0_1}$, 
$m_{\tilde{\chi}^0_2}$, $\sigma_L\{12\}$ and $\sigma_R\{12\}$ 
for a given value of the mass parameter $M_1$ are displayed with respect 
to their measured values:
\begin{eqnarray}
{\rm Ratio}=m^{th}_{\tilde{\chi}^0_i}(M_1)/m^{meas}_{\tilde{\chi}^0_i}
\qquad {\rm and} \qquad 
\sigma^{th}(M_1)/\sigma^{meas}
\end{eqnarray}
 In the left panel the curves  all meet
in exactly one point  proving that 
\begin{eqnarray}
{\sf P1}: \qquad M_1=100.5\, {\rm GeV}
\end{eqnarray}
is the correct solution. Additional consistency
checks  can be provided by measuring the production
cross sections $ \sigma_T\{12\}$, if transversely
polarized electron and positron beams are available.

\subsection{The supersymmetric Yukawa couplings} 

The identity of the  SUSY Yukawa couplings $g_{_{\tilde{W}}}$ and 
$g_{_{\!\tilde{B}}}$ with  the SU(2) and U(1) gauge couplings $g$ and $g'$, 
which is of fundamental importance in supersymmetric theories, can be 
tested very accurately in neutralino pair--production. This 
analysis is one of the final targets of LC experiments which should provide
a complete picture of the electroweak gaugino sector with resolution at 
least at the per-cent level.

We assume here that the SU(2) gaugino/higgsino parameters in the
CP--invariant theory have been
pre--determined in the chargino sector and the U(1) parameter $M_1$
has been extracted from the neutralino mass spectrum.
The equality between the Yukawa and the gauge couplings can be tested
precisely by making use of electron (and positron) beam
polarization.  Varying the left--handed and right--handed Yukawa
couplings leads to a significant change in the corresponding left--handed
\begin{figure}[hbt]
\begin{center}
 \epsfig{file=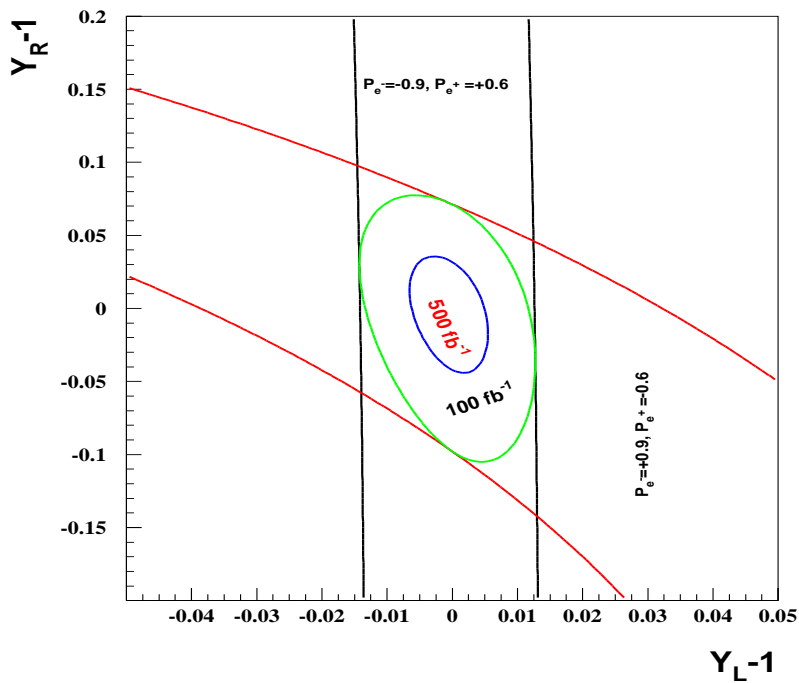, width=12cm, height=10cm}
\caption{\it Contours of the cross sections $\sigma_L\{12\}$ and 
       $\sigma_R\{12\}$ in the plane of the Yukawa couplings $g_{_{\tilde{W}}}$
       and $g_{_{\!\tilde{B}}}$ normalized to the SU(2) and U(1) gauge
       couplings $g$ and $g'$ $\{Y_L=g_{{\tilde W}}/g,\, 
       Y_R=g_{{\tilde B}}/g'\,\}$ for the set ${\sf RP1}$ at the $e^+e^-$ 
       c.m. energy of 500 GeV; the contours correspond to the integrated 
       luminosities 100 and 500 fb$^{-1}$ and the longitudinal polarization 
       of electron and positron beams of 90\% and 60\%, respectively.}
\label{fig:pol}
\end{center}
\end{figure}
and right--handed production cross sections. Combining the
measurements of $\sigma_R$ and $\sigma_L$ for the process
$e^+e^-\rightarrow \tilde{\chi}^0_1\, \tilde{\chi}^0_2$ process, the
Yukawa couplings $g_{_{\tilde{W}}}$ and $g_{_{\!\tilde{B}}}$ can be
determined to quite a high precision as demonstrated in
Fig.~\ref{fig:pol}.  The $1\sigma$ statistical errors have been derived
for an  integrated luminosity of $\int {\cal L}\, dt =100$ and $500$
fb$^{-1}$ and for partially polarized beams. 

Combined with the measurement of the $\tilde{W}e\tilde{\nu}$ Yukawa 
coupling, including the analysis of angular distributions, in the chargino 
sector, it is possible to check the
crucial SUSY relation between the gauge couplings and the supersymmetric
Yukawa couplings in a comprehensive way.

\subsection{ The complete MSSM neutralino system}

The measurements of the chargino--pair production processes
$e^+e^-\rightarrow\tilde{\chi}^+_i\tilde{\chi}^-_j$ ($i,j$=1,2)
carried out with polarized beams can be used for a complete
determination of the basic SUSY parameters $\{M_2,\,|\mu|,
\, \Phi_\mu\, ;\tan\beta\}$ 
in the chargino sector with high precision\footnote{
  The sine of the phase $\Phi_\mu$ can be determined by measuring the
  sign of observables associated with the normal
  $\tilde{\chi}^\pm_{1,2}$ polarizations \cite{CDGKSZ}.}.  In this
section, it will be demonstrated analytically in the general
CP--noninvariant theory that the real and
imaginary parts of 
the U(1) gaugino mass $M_1$ can be determined  subsequently from the
measurements of (i) either three neutralino masses or/and (ii) 
from the masses of two light neutralinos and one neutralino--pair
production cross section such as $\sigma\{12\}$.

Each of the four invariants $a$, $b$, $c$,
$d$ of the matrix
${\cal M}{\cal M}^\dagger $, defined in eq.~(\ref{eq:inv}), is a
second--order polynomial of  
$\real{M_1}=|M_1|\,\cos\Phi_1$ and $\imag{M_1}=|M_1|\,\sin\Phi_1$.
Therefore, each of the characteristic equations in the set 
(\ref{eq:characteristic}) for the  neutralino mass squared  
can be cast  into the form
\begin{figure}[tb]
\hskip -1cm
 \epsfig{file=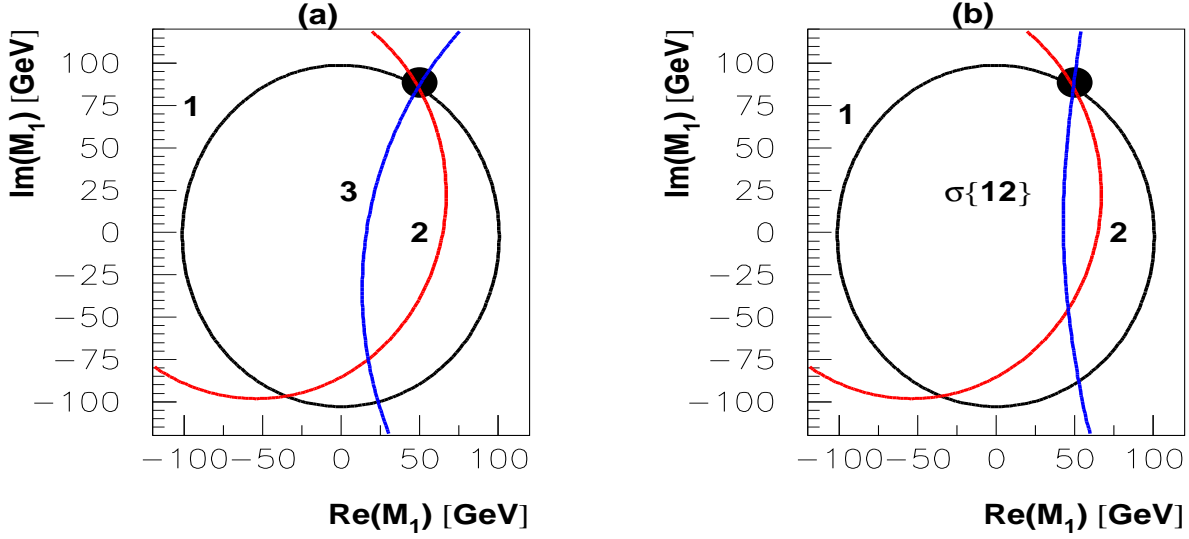, width=19cm, height=9cm}
\caption{\it The contours of (a) three measured neutralino masses 
$m_{\tilde{\chi}^0_i}$ ($i=1,2,3$), and
(b) two neutralino masses (1,2) and one neutralino production cross section
$\sigma_{tot}\{12\}$ in the $\{\real{M_1},\, \imag{M_1}\}$ plane;
the parameter set ${\sf RP1''}$ $\{\tan\beta=10, \, M_2=190.8\, {\rm GeV},\,
|\mu|=365.1\,{\rm GeV}, \, \Phi_\mu=\pi/4\}$ is taken from the
chargino sector.}
\label{fig:contour}
\end{figure}
\begin{figure}[htb]
\hskip -1cm
 \epsfig{file=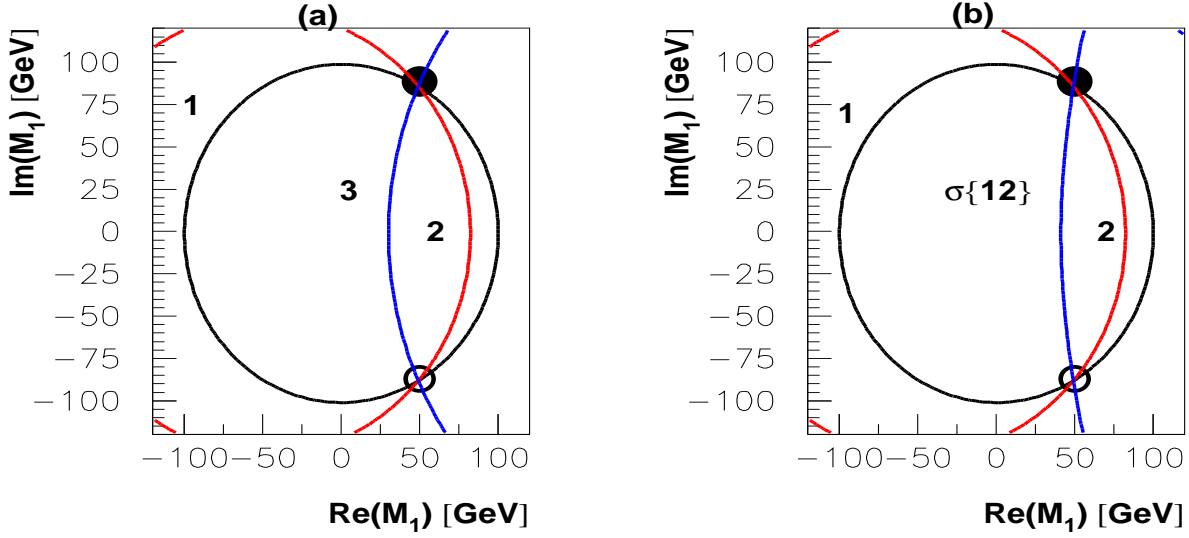, width=19cm, height=9cm}
\caption{\it The contours of (a) three measured neutralino masses 
and (b) two measured light neutralino masses and 
one neutralino production cross section,
$\sigma_{tot}\{12\}$ in the $\{\real{M_1},\, \imag{M_1}\}$ plane for
the CP--violating case  ${\sf RP1'}$:
$\{\tan\beta=10,  \, M_2=190.8\,
{\rm GeV}, \mu=365.1\,{\rm GeV}, \, \sin\Phi_\mu=0\}$. }
\label{fig:MwithCP}
\end{figure}
\begin{eqnarray}
(\real{M_1})^2+(\imag{M_1})^2+ u_i\real{M_1}+ v_i\imag{M_1}
 = w_i \qquad (i=1,2,3,4)
\label{eq:Mphase}
\end{eqnarray}
The coefficients $u_i$, $v_i$ and $w_i$ are functions of the 
parameters $\tan\beta$, $M_2$, $|\mu|$, $\Phi_\mu$ pre--determined in
the chargino sector, and the mass
$m^2_{\tilde{\chi}^0_i}$;  the coefficient $v_i$ is necessarily
proportional to $\sin\Phi_\mu$ because physical masses are CP--even.
For each neutralino mass, eq.~(\ref{eq:Mphase}) defines a circle
in the $\{\real{M_1},\imag{M_1}\}$ plane. As a result, the measurement
of three  neutralino masses leads to an unambiguous
determination of the modulus and the phase of $M_1$,
cf. Fig.~\ref{fig:contour}(a).   With only two
light neutralino masses, the two--fold ambiguity can be resolved 
by exploiting the measured cross section $\sigma\{12\}$, as shown in
Fig.~\ref{fig:contour}(b).  However, if the phase $\sin\Phi_\mu$ vanishes,
there remains a two--fold discrete sign ambiguity in $\imag{M_1}$, as
demonstrated  in Fig.\ref{fig:MwithCP}.

\section{Closure of the neutralino system}

Since the reconstruction of the mass and mixing parameters is easy if
all four neutralino states are detected, stringent tests of the
four--state closure can be designed. Models with additional chiral or
vector superfields, for example, give rise to extensions of the neutralino
sector in general.

The four--state mixing of neutralinos in the minimal supersymmetric
extension of the Standard Model induces sum rules for the neutralino
couplings. 
They can be formulated in terms of the squares of the bilinear
charges, {\it i.e.} the factorized elements of the quartic charges.
This  follows
from the unitarity of the diagonalization matrices. If all possible
neutralino states are summed up, the following general sum rules can
be derived at tree level:
\begin{eqnarray}
&&   \sum_{i,j=1}^4\, {\cal Z}_{ij} {\cal Z}^*_{ij} = \frac{1}{2} 
\hskip 2cm 
   \sum_{i,j=1}^4\, g_{Lij} g^*_{Lij} = \frac{1}{16 c^4_W s^4_W}
\nonumber \\
&&  \sum_{i,j=1}^4\, {\cal Z}_{ij} g^*_{Lij}=0 
\hskip 2cm
  \sum_{i,j=1}^4\, g_{Lij} g^*_{Rij} = \frac{1}{4c^4_W} \nonumber \\
&&  \sum_{i,j=1}^4\, {\cal Z}_{ij} g^*_{Rij} =0 \hskip 2cm
  \sum_{i,j=1}^4\, g_{Rij} g^*_{Rij} = \frac{1}{c^4_W} 
\label{eq:sum rule} 
\end{eqnarray}
The right--hand side of the sum rules is {\it independent of the parameters
in the neutralino system} and it is given solely by the gauge group.
Therefore, 
evaluating these sum rules experimentally, it can be tested whether
the four--neutralino system $\{\tilde{\chi}_1^0, \tilde{\chi}_2^0,
\tilde{\chi}_3^0,\tilde{\chi}_4^0\}$ forms a closed system, or whether
additional states at high mass scales mix in, signaling the existence
of an extended gaugino system.

The validity of the sum rules is reflected in both the quartic 
charges and the production cross sections. However, due to mass effects
and the $t$-- and $u$--channel selectron exchanges, it is not 
straightforward to derive the sum rules for the quartic charges and 
the production cross sections in practice. {\it Asymptotically} 
at high energies, however, the sum rules in eq.~(\ref{eq:sum rule})  
can be transformed 
directly into sum rules for the associated cross sections:
\begin{eqnarray}
\lim_{s\rightarrow \infty}\,s\,\sum_{i\leq
     j}^4\,\sigma\{ij\} 
     =\frac{\pi\alpha^2}{48\,c^4_W s^4_W}\left[64 s^4_W-8 s^2_W+ 5\right]
\end{eqnarray}
The approach to the asymptotic form of the sum rules depends on the mass
parameters of the theory. (The mixing parameters, weighted by the
physical neutralino masses, can be summed up to polynomials of the gaugino and
higgsino mass parameters, as demonstrated in the appendix.)

In Fig.~\ref{fig:sr} the exact values for
the summed-up cross sections normalized to the asymptotic value are
shown  for the
reference point ${\sf RP1}$. The final
state $\tilde{\chi}^0_1\tilde{\chi}^0_1$ is  invisible in
$R$-parity invariant theories, and its detection is
difficult. Nevertheless, it can be studied directly  by
photon  tagging in the final state 
$ \gamma \tilde{\chi}^0_1\tilde{\chi}^0_1$, which can be observed at
the LC. Indirectly the $\tilde{\chi}^0_1\tilde{\chi}^0_1$ cross
section can be predicted by extracting, hypothetically, the MSSM
parameters from the observed cross sections. The subsequent 
failure of saturating the sum rules would then be sufficient to conclude 
that the neutralino system of the MSSM is not closed indeed 
and additional states mix in.

\begin{figure}[tb]
\begin{center}
 \epsfxsize=10cm \epsfbox{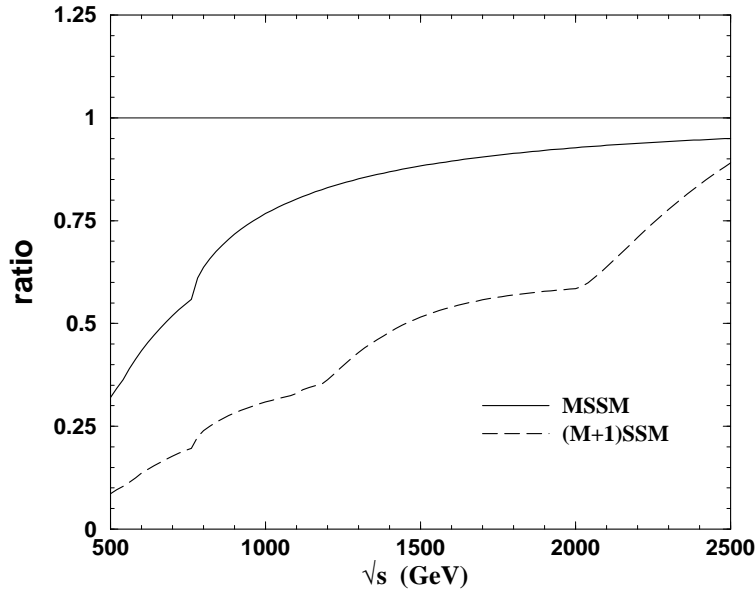}
\caption{\it The energy dependence of the sum of all the 
  neutralino--pair production cross sections normalized to the
  asymptotic form of the summed up cross section; the solid line
  represents the exact sum in the MSSM;
   the dashed line the sum of the cross sections 
  for the first four neutralino states in a specific parameter set 
  of the (M+1)SSM. }
\label{fig:sr}
\end{center}
\end{figure}

More specifically, extended SUSY models with $n$ SU(2)
doublet\footnote{An even number of doublets is needed to cancel the
  chiral anomaly properly.} and $m$ SU(2) singlet chiral superfields
may be considered in general.  In these extended models,
diagonalization of the mass matrix leads to  
$(2+2n+m)$ neutralino mass eigenstates. 
The fermion (higgsino) components of
the chiral fields do not modify the structure of the $\tilde{\chi}^0_i
e\tilde{e}_{L,R}$ vertices. While the higgsino singlets do not change the
structure of the $Z$--neutralino--neutralino vertices, the neutral
component of each additional higgsino doublet with  hypercharge $\pm
1/2$ couples to the $Z$ boson exactly in the same way as
$\tilde{H}^0_{1,2}$. So, the $Z$--neutralino--neutralino couplings are
modified to read
\begin{eqnarray}
&& \langle\tilde{\chi}^0_{iL}\, |Z|\tilde{\chi}^0_{jL}\rangle 
  = -\frac{g}{2\,c_W} \sum_{a=1}^n
     \left[N_{i(1+2a)}N^{*}_{j(1+2a)}-N_{i(2+2a)}N^{*}_{j(2+2a)}\right]
   \nonumber\\
&& \langle\tilde{\chi}^0_{iR}\,|Z|\tilde{\chi}^0_{jR}\rangle 
  = +\frac{g}{2\,c_W} \sum_{a=1}^n
     \left[N^{*}_{i(1+2a)}N_{j(1+2a)}-N^{*}_{i(2+2a)}N_{j(2+2a)}\right]
\end{eqnarray}
The sum rule, following
from the unitarity of the $(2+2n+m)\times (2+2n+m)$
mixing matrix, for the 
pair--production cross sections of {\it all} states is extended to 
\begin{eqnarray}
\lim_{s\rightarrow \infty}\,s\,\sum_{i\leq j}^{2+2n+m}\, 
          \sigma \{ij\} = 
          \frac{\pi\alpha^2}{48\, c^4_W s^4_W}
             \left[\, 2\,(8 s^4_W-4 s^2_W+ 1)\, n+ 48 s^4_W + 3\right]
\label{eq:srext}
\end{eqnarray}
The right--hand side of eq.~(\ref{eq:srext}) is independent of the 
number $m$ of higgsino
singlets and it reduces to the sum rule in the
MSSM for $n=1$.

A typical example is provided by  the extended (M+1)SSM scenario  which
incorporates an additional gauge singlet superfield \cite{NMSSM}, but
does not change the structure of the charged sector.
The superpotential of the
(M+1)SSM is given by
\begin{eqnarray}
W_{(M+1)SSM} = W_Y + \lambda\, S H_1 H_2 +\frac{1}{3}\kappa S^3
\end{eqnarray}
where $W_Y$ accounts for the lepton and quark Yukawa interactions. In
this model, an effective $\mu=\lambda s$ term is generated when the
scalar component of the singlet $S$ acquires a vacuum expectation
value $s=\langle S \rangle $.  The fermion component of the
singlet superfield (singlino) will mix with neutral gauginos and
higgsinos after electroweak gauge symmetry breaking, changing the
neutralino mass matrix to the 5$\times$5 form
\begin{eqnarray*}
{\cal M}_{(M+1)SSM}=
\left(\begin{array}{ccccc}
|M_1|\,{\rm e}^{i\Phi_1} & 0 & -m_Z c_\beta s_W & m_Z s_\beta s_W & 0 
\\[2mm]
   0    &   M_2      &  m_Z c_\beta c_W  & -m_Z s_\beta c_W & 0 \\[2mm]
-m_Z c_\beta s_W & m_Z c_\beta c_W & 0 & -|\mu|\, {\rm e}^{i\Phi_\mu} & 
     -|M_\lambda|\,s_\beta\, {\rm e}^{i\Phi_\lambda} \\[2mm]
 m_Z s_\beta s_W &-m_Z s_\beta c_W & -|\mu|\, {\rm e}^{i\Phi_\mu} & 0 &
    -|M_\lambda|\,c_\beta\, {\rm e}^{i\Phi_\lambda} \\[2mm]
  0 & 0 & -|M_\lambda|\,s_\beta\, {\rm e}^{i\Phi_\lambda} &
     -|M_\lambda|\,c_\beta\, {\rm e}^{i\Phi_\lambda} &
     \,2 |M_\kappa|\, {\rm e}^{i\Phi_\kappa}
                  \end{array}\right)\
\label{eq:NMSSM mass matrix}
\end{eqnarray*}
where $|M_\lambda|\, {\rm e}^{i\Phi_\lambda}\equiv \lambda v$ and
$|M_\kappa|\, {\rm e}^{i\Phi_\kappa}\equiv \kappa s$.

In some regions of the parameter space \cite{Franke} the singlino may
be the lightest supersymmetric particle, weakly mixing with other
states. Then displaced vertices in the (M+1)SSM may be generated,
which would signal the extension of the minimal model.  If the
spectrum of the  four lighter neutralinos in the extended model is similar
to the spectrum in the MSSM but the mixing is substantial, 
discriminating the models by analyzing the mass spectrum becomes very
difficult.   Studying in this
case the summed-up cross sections of the four light neutralinos may
then be a crucial method to reveal the structure of the neutralino system.

In Fig.\ref{fig:sr} the exact sum rules are also included for a
possible scenario of the (M+1)SSM; the parameters $M_1=1000$ GeV,
$M_2=169$ GeV, $\mu=-263$ GeV, $\tan\beta=10$ and $M_\lambda=263$ GeV,
$M_\kappa=-59$ GeV, give rise to one very heavy neutralino with 
$m_{\tilde{\chi}^0_5}\sim 1000$ GeV, and to four lighter
neutralinos with masses within 2 -- 5 GeV equal to the neutralino
masses for  the
${\sf RP1}$ point of the MSSM.  Due to the incompleteness of these states
below the thresholds for producing the heavy neutralino, the (M+1)SSM value
differs significantly from the corresponding sum rule of the MSSM.
Therefore, even if the extended neutralino states are very heavy,
the study of sum rules can shed light on the underlying structure of the
supersymmetric model.
\vskip 7mm

\noindent {\bf Addendum: Charginos}\\[3mm]
{\it Asymptotically} at high energies 
the sum rule 
\begin{eqnarray}
\lim_{s\rightarrow \infty}\,s\,\sum_{i j}^2\,\sigma^{\pm}\{ij\}
     =\frac{\pi\alpha^2}{24\,c^4_W s^4_W}\left[8 s^4_W-8 s^2_W+ 5\right]
\end{eqnarray}
for the summed-up chargino cross sections \cite{CDGKSZ} can be derived
in the same way. In analogy to the neutralino system, the approach to
asymptotia depends on the gaugino and higgsino parameters, cf. appendix.

\section{Conclusions}

In the first part of this analysis we have derived  the
mass eigenvalues  and the mixing matrix of the MSSM neutralino system
including CP violation. The problem  has been solved analytically, and
a compact representation has been found in the limit of large  SUSY
gaugino and higgsino mass parameters compared to the scale of
electroweak symmetry breaking. Unitarity quadrangles  have been
introduced,  distinctly
different from CKM and MNS polygons due to the Majorana nature of the
neutralinos. They illustrate nicely the specific
realization of CP violation through the two distinct sets of phases in the
system. In this way the solution of the MSSM neutralino system has been
advanced to a level analogous to the chargino system.

If the chargino system is solved for the SU(2) 
parameters $\{M_2,|\mu|,\Phi_\mu; \tan\beta\}$, the neutralino mass
spectrum is sufficient to extract the U(1) gaugino mass parameter 
$\{|M_1|,\Phi_1\}$.  Three (light)
neutralino masses $m_{\tilde{\chi}^0_{1,2,3}}$ or/and two light
neutralino masses $m_{\tilde{\chi}^0_{1,2}}$ supplemented by the
production cross section $\sigma\{12\}$ for the neutralino pair
$\tilde{\chi}^0_1 \tilde{\chi}^0_2$, allow us to extract $\{|M_1|,
\Phi_1\}$ unambiguously, and with a two--fold ambiguity for the sign
of $\sin\Phi_1$ if $\sin\Phi_\mu$ vanishes. This discrete ambiguity
can be solved by measuring the normal neutralino  polarization
and/or the cross section $\sigma_N$ with initial transverse 
beam polarization. 
All fundamental SU(2)$\times$U(1) gaugino and higgsino
parameters can therefore be derived analytically in the combined
chargino $\oplus$ neutralino system from measured mass and mixing
parameters.

Sum rules for the production cross sections can be used at high energies
to probe whether the four--state neutralino system is closed
or  whether
additional states mix in from potentially very high scales.\\

{\it To summarize.} The measurement of the processes
$e^+e^-\rightarrow \tilde{\chi}^0_i \tilde{\chi}^0_j$ ($i,j$=1,2,3,4),
carried out with polarized beams and combined with the analysis of the
chargino system $e^+e^-\rightarrow \tilde{\chi}^+_i \tilde{\chi}^-_j$
($i,j$=1,2), can be used to perform a complete and precise analysis of
the basic 
SUSY parameters in the gaugino/higgsino sector $\{M_1, M_2, \mu;
\tan\beta\}$.  The
chargino/neutralino system of the MSSM at tree level is therefore
under analytical 
control {\it in toto}.

Since the analysis can be performed with high precision, this set
provides a solid platform for extrapolations to scales eventually near
the Planck scale where the fundamental supersymmetric theory may be
defined.\\

\appendix

\section*{APPENDIX}

\section{Sum Rules: Approach to Asymptotia}

While the sum rules in the asymptotic limit do not depend on any supersymmetry
parameters of the gaugino/higgsino sector but only on the gauge group, 
the {\it approach} to asymptotia involves the neutralino and chargino 
masses. Nevertheless, the
sums of the mixing parameters weighted by these masses, can be expressed by
the fundamental gaugino and higgsino mass parameters in closed form.\\

\subsection{Neutralino system}

The following mass weighted sum rules\footnote{We introduce the abbreviations
$s_{2W}=\sin 2\theta_W$ and $c_{2W}=\cos 2\theta_W$.}
\begin{eqnarray}
&& \sum^4_{ij} m^2_i|{\cal Z}_{ij}|^2 =\frac{m_Z^2+2|\mu|^2}{4}\qquad
   \hskip 1cm 
   \sum^4_{ij} m_i m_j |{\cal Z}_{ij}|^2 =-\frac{|\mu|^2}{2}\nonumber\\
&& \sum^4_{ij} m^2_i|g_{Rij}|^2 =\frac{|M_1|^2+m^2_Z s^2_W}{c^4_W} \quad
   \hskip 0.7cm
   \sum^4_{ij} m^2_i|g_{Lij}|^2 =\frac{|M_1|^2s^2_W+|M_2|^2 c^2_W
                     +m^2_Z c^2_{2W}}{16\,c^4_W s^4_W}\nonumber\\
&& \sum^4_{ij} m_i m_j g_{Rij}^2 = \frac{|M_1|^2}{c^4_W} \  \
   \hskip 2.45cm
   \sum^4_{ij} m_i m_j g_{Lij}^2 = \frac{|M_1 s^2_W+ M_2
                     c^2_W|^2}{16c^4_W s^4_W}\nonumber\\
&& \sum^4_{ij} m_i m_i {\cal Z}_{ij} g_{Rij} 
       = \frac{m^2_Z s^2_W\,c_{2\beta}}{2 c^2_W}\qquad 
   \hskip 0.63cm
   \sum^4_{ij} m_i m_i {\cal Z}_{ij} g_{Lij} 
       = \frac{m^2_Z\, c^2_{2W}c_{2\beta}}{8 c^2_W s^2_W}
\end{eqnarray}
and
\begin{eqnarray}
\sum^4_{ij} m^2_i m^2_j |g_{Lij}|^2\!\! &=& \left[|M_1|^2s^2_W+|M_2|^2 c^2_W
   + m^2_Z c^2_{2W}\right]^2/16\, c^4_W s^4_W\nonumber\\
 \sum^4_{ij} m^2_i m^2_j |g_{Rij}|^2\!\! &=& 
     \left[|M_1|^2+ m^2_Zs^2_W\right]^2/c^4_W
\end{eqnarray}
can be used in the sum of the neutralino cross sections
\begin{eqnarray}
\lim_{s\rightarrow \infty}\,s\,\sum_{i\leq j}^4\,\sigma\{ij\} 
     =\frac{\pi\alpha^2}{48\,c^4_W s^4_W}\left\{\, [\,64 s^4_W-8
      s^2_W+ 5\,]\, +\Delta^0_1/s+\Delta^0_2/s\right\}
\label{eq:nsub-leading}
\end{eqnarray}
to calculate the coefficients $\Delta^0_1$ and $\Delta^0_2$ which control
the approach to asymptotia:
\begin{eqnarray}
\Delta^0_1 &= & (8 s^4_W- 4 s^2_W+1) m^2_Z + 3 m^2_{\tilde{e}_L}
                + 48 s^4_W m^2_{\tilde{e}_R}\nonumber\\
	   && -192 s^4_W (|M_1|^2+m^2_Z s^2_W)
	      -12(|M_1|^2 s^2_W+|M_2|^2 c^2_W + m^2_Z c^2_{2W})\nonumber\\
	   && +6\left\{ |M_1-M_2|^2 s^2_{2W}/4 
	      +m^2_Z c^2_{2W} (1+c_{2W} c_{2\beta})-m^2_{\tilde{e}_L}\right\}
	       \log\left(1+s/m^2_{\tilde{e}_L}\right)\nonumber\\
	   && +48 s^4_W\left\{ m^2_Z s^2_W (2+c_{2\beta})- 2m^2_{\tilde{e}_R}
	      \right\}\log\left(1+s/m^2_{\tilde{e}_R}\right)
	       \nonumber\\ 
\Delta^0_2 &=&\frac{3}{m^2_{\tilde{e}_L}}
      \left\{|M_1|^2 s^2_W +|M_2|^2 c^2_W+m^2_Z c^2_{2W}\right\}^2
      + \frac{48 s^4_W}{m^2_{\tilde{e}_R}}\left\{|M_1|^2+m^2_Z s^2_W\right\}^2
\label{eq:delta-zero}
\end{eqnarray}
The approach to asymptotia is fast for the reference point chosen before.
For $\sqrt{s}=2$ TeV the form including the subleading terms in eqs.
(\ref{eq:nsub-leading}) and (\ref{eq:delta-zero}) has reached already 90 percent
of the asymptotic limit.

\subsection{Chargino system}

The coefficients $\Delta^\pm_{1,2}$ in the sum rule for the chargino cross
sections 
\begin{eqnarray}
\lim_{s\rightarrow \infty}\,s\,\sum_{ij}^2\,\sigma^\pm\{ij\} 
     =\frac{\pi\alpha^2}{24\,c^4_W s^4_W}\left\{\, [\,8 s^4_W-8
      s^2_W+ 5\,]\, +\Delta^\pm_1/s+\Delta^\pm_2/s\right\}
\label{eq:csub-leading}
\end{eqnarray}
can be evaluated in the same way:
\begin{eqnarray}
\Delta^\pm_1 &=& 2(6 s^6_W+5 s^4_W-8 s^2_W+2)\, m^2_Z
	    - 3 (8 s^4_W- 4 s^2_W +1) \, m^2_W \nonumber\\ 
	 && + 18\, c^4_W m^2_{\tilde{\nu}}-24 c^4_W (|M_2|^2+ 2m^2_W c^2_\beta) 
	     \nonumber\\
	 && -12 c^2_W \left[m^2_{\tilde{\nu}}\, c^2_W
	               + m^2_W c^2_\beta (2s^2_W-1) \right]
	 \log\left(1+s/m^2_{\tilde{\nu}}\right)\nonumber\\[2mm]
\Delta^\pm_2 &=& \frac{6}{m^2_{\tilde{\nu}}}\, c^4_W 
                 (|M_2|^2 + 2m^2_W c^2_\beta)^2
\end{eqnarray}
Again, the approach to asymptotia is fast for the parameter set under
discussion.\\

\subsubsection*{Acknowledgments}

SYC was supported by the Korea Research Foundation Grant
(KRF--2000--015--DS0009), JK by the KBN Grant No. 
2P03B 060 18. 
The work was supported in part by the European Commission 5-th
Framework Contract HPRN-CT-2000-00149. 
SYC and JK acknowledge the hospitality extended to them at DESY by
Profs.\ Klanner and Wagner. We also thank H. Fraas, H. Haber, 
T. Han, W. Hollik, J.-L. Kneur,
U. Martyn and G. Moultaka 
for useful  discussions.


%

%

\end{document}